\documentclass[journal]{IEEEtran}
\usepackage[utf8]{inputenc}
\usepackage[pdftex]{graphicx}
\usepackage{epstopdf}
\usepackage[utf8]{inputenc} % allow utf-8 input
\usepackage[T1]{fontenc}    % use 8-bit T1 fonts
\usepackage{hyperref}       % hyperlinks
\usepackage{url}            % simple URL typesetting
\usepackage{booktabs}       % professional-quality tables
\usepackage{amsfonts}       % blackboard math symbols
\usepackage{nicefrac}       % compact symbols for 1/2, etc.
\usepackage{microtype}      % microtypography
\newenvironment{proof}[1][Proof]{{\it #1. } }{\ \rule{0.5em}{0.5em}}
\usepackage{graphics, theorem, times, cite}

\usepackage{tikz}

\usepackage{bbm}
\usepackage{mathrsfs}
\usepackage{adjustbox}
\usepackage{makecell}
\usepackage{wrapfig}

\usepackage{amssymb,amsmath,amsfonts,bm,epsfig,theorem}
\usepackage{rotating,setspace,latexsym,epsf,color,dsfont,caption,mathtools}
\usepackage{subfigure}

\usepackage{algorithm}
\usepackage{algorithmic}
\usepackage[shortlabels]{enumitem}

\input{mysymbol.sty}
\usepackage{needspace}

\newcommand{\QED}{\hfill\ensuremath{\blacksquare}}

\def\E{\E}

\newtheorem{assumption}{\hspace{0pt}\bf Assumption}

\newtheorem{proposition}{\hspace{0pt}\bf Proposition}

\newtheorem{theorem}{\hspace{0pt}\bf Theorem}

\newtheorem{remark}{\hspace{0pt}\bf Remark}

\title{Learning Decentralized Wireless Resource Allocations with Graph Neural Networks}
\author{Zhiyang Wang \quad Mark Eisen \quad Alejandro Ribeiro\thanks{Supported by Intel Science and Technology Center for Wireless Autonomous Systems and ARL DCIST CRA W9111NF-17-2-0181. Z. Wang and A. Ribeiro are with the Department of Electrical and Systems Engineering, University of Pennsylvania, PA, email: \{zhiyangw, aribeiro\}@seas.upenn.edu. M. Eisen is with Intel Corporation, Hillsboro, OR, email: mark.eisen@intel.com. Preliminary results presented in \cite{wang2020decentralized, wang2021unsupervised}. } }
\begin{document}
\maketitle
\begin{abstract}
We consider the broad class of decentralized optimal resource allocation problems in wireless networks, which can be formulated as a constrained statistical learning problems with a localized information structure. We develop the use of Aggregation Graph Neural Networks (Agg-GNNs), which process a sequence of delayed and potentially asynchronous graph aggregated state information obtained locally at each transmitter from multi-hop neighbors. We further utilize model-free primal-dual learning methods to optimize performance subject to constraints in the presence of delay and asynchrony inherent to decentralized networks. We demonstrate a permutation equivariance property of the resulting resource allocation policy that can be shown to facilitate transference to dynamic network configurations. The proposed framework is validated with numerical simulations that exhibit superior performance to baseline strategies.
\end{abstract}

\begin{IEEEkeywords}
Resource allocation, decentralized, graph neural networks, deep learning 
\end{IEEEkeywords}

%%%%%%%%%%%%%%%%%%%%%%%%%%%%%%%%%%%%%%%%%%%%%%%%%%%%%%%%%%%%%%%%%%%%%
%%%%%%%%%%%%%%%%%%%%%%%%%%%%%%%%%%%%%%%%%%%%%%%%%%%%%%%%%%%%%%%%%%%%
%%%   S  E  C  T  I  O  N   %%%%%%%%%%%%%%%%%%%%%%%%%%%%%%%%%%%%%%
%%%%%%%%%%%%%%%%%%%%%%%%%%%%%%%%%%%%%%%%%%%%%%%%%%%%%%%%%%%%%%%%%%%%
%%%%%%%%%%%%%%%%%%%%%%%%%%%%%%%%%%%%%%%%%%%%%%%%%%%%%%%%%%%%%%%%%%%%%
\section{Introduction}
\label{sec:intro}

Rapid growth of user demand and number of access devices strains the ability of wireless systems to meet quality of service requirements. This challenge calls for the use of optimal resource allocation policies that make the best possible use of available bandwidth and power resources \cite{Ribeiro2012}. Optimal policies are, however, intractable to find in all but the simplest scenarios. In practice, heuristics that try to approximate optimal policies are used instead {in both convex and non-convex conditions}; see, e.g., \cite{zhang2006stochastic, wang2016dynamic, khalili2020joint, chen2011round, sangaiah2020iot, liu2019stochastic, tun2020joint}. Recent years have seen increasingly successful attempts at using \emph{learned} heuristics where a model is trained to find good resource allocations. Learned heuristics can often outperform designed heuristics but they also have some other advantages. They are computationally less costly \cite{xu2019energy, sun2017learning, sun2018learning}. They can learn from interactions with the environment and are therefore not necessarily reliant on access to channel and rate models \cite{xu2017deep,lee2018deep,eisen2019learning}. And, if the parametrization is suitably chosen, they can scale to networks with large numbers of transceivers \cite{cui2019spatial, lee2019graph,eisen2020optimal, naderializadeh2020wireless, guo2020structure, shen2020graph, chowdhury2021unfolding, chowdhury2021efficient, jiang2021learning, hu2020iterative,hu2021joint }. In this paper we explore a fourth potential advantage of learned heuristics: The feasibility of distributed implementations.

In distributed resource allocation policies transceivers have knowledge of their local radio environment only and make local decisions based on this information. Distributed policies have been recognized as a necessity since the early days of power control \cite[Ch. 6]{Viterbi95} given the rapid channel changes that are characteristics of wireless communications. In their simplest incarnations, distributed policies map local information to individual node decisions; see e.g, \cite{chavez1997challenger, belleschi2011performance}. In their more involved versions, distributed policies exchange information with neighboring nodes and base their decisions on an augmented space that includes their observations and the messages they receive from neighbors \cite{shi2011iteratively}. 

This evolution from leveraging purely local information towards leveraging information from neighboring agents can also be seen in learned policies. The development of data-driven policies that exploit local information \cite{challita2017proactive, li2018intelligent, naparstek2018deep, kim2018completely, de2018team} has led to the development of data-driven policies that further incorporate information from nearby agents \cite{ye2019deep, zhao2019deep, nasir2019multi, naderializadeh2020resource}. Existing works on decentralized resource allocation, however, largely ignore the global structure of the wireless system beyond nodes' immediate neighbors. They, moreover, do not address the inevitability of information delay and asynchrony between devices. These environmental factors motivate a learning approach that can tolerate delays, operate without synchrony, and
leverage information beyond a node's immediate neighborhood. The main contribution of this paper is to develop techniques for learning decentralized policies with these three characteristics. 

Specifically, we develop a scalable and learning-based approach for solving a broad class of decentralized resource allocation problems in which a global network utility is optimized subject to system constraints. We leverage the tools of unsupervised learning to design localized resource policies that optimize performance and adhere to given constraints in realistic decentralized environments subject to delay in information exchanges and asynchronous working clocks. In particular, we propose the use of Aggregation Graph Neural Networks (Agg-GNNs) \cite{gama2019convolutional}, which utilize successive information exchanges between neighboring nodes to allow devices to locally accumulate global network state information.{ The multiple layers of processing delayed information after signal aggregations in Agg-GNNs allow the gathering of correlated spatial and temporal information of the global wireless network, which is also the main distinction between general graph neural networks \cite{gama2019convolutional} and our Agg-GNN method.} In Agg-GNNs, graph shift operations capture the asynchronous states of communication links and device nodes of the {global} network in a manner that is invariant to permutations of the network. Because the local policy is common among network nodes, the proposed architecture can be implemented in a manner that is invariant to network size and thus permitting a transference to larger networks. This is indeed of critical importance in learning applications of wireless systems, where only fixed networks of limited size may be available at the time of training but systems may change frequently during execution. {We note the recent use of alternative graph neural network architectures in approximating distributed resource allocation policies \cite{shen2020graph, chowdhury2021unfolding, chowdhury2021efficient} in a stricter class of problems, disregarding potential existence of constraints, delay, and asynchrony.} {The architecture of Agg-GNN has been successfully implemented in other scenarios, e.g. robot swarming contrl \cite{tolstaya2020learning} while here we consider a novel problem setting in wireless communication networks. Different from our previous works \cite{wang2020decentralized}\cite{wang2021unsupervised}, we make an extension to consider a general decentralized, asynchronous resource allocation scenario and we further consider a more practical correlated channel model.}

The proposed Agg-GNN architecture contains important structural properties that allow for decentralized implementation and network transference, though the filter weights must be carefully trained to optimize performance and satisfy constraints. We utilize an unsupervised, model-free method that can optimize generic resource allocation problems subject to the environmental limitations of decentralized network architectures. Our main contributions are as follows:

\begin{itemize}
    \item We introduce the Aggregation Graph Neural Networks (Agg-GNNs) to parameterize a local decentralized policy for general constrained resource allocation problems. The Agg-GNN successively aggregates global network information at each node, either through synchronous or asynchronous communications.
    \item We establish the permutation equivariance of the optimal Agg-GNN resource allocation policy, which facilitates transference of the learned Agg-GNN to other networks of varying size.
    \item We {adapt the primal-dual learning method to operate in the an asynchronous and model-free manner} for training Agg-GNNs to solve constrained resource allocation problems without explicit model knowledge.
    \item We perform extensive numerical analysis of the performance and transference of the Agg-GNN in a classical power allocation problem, in which we demonstrate that superior performance of the proposed framework relative to existing baselines.
\end{itemize}

The rest of this paper is organized as follows. In Section \ref{sec:prob}, we introduce the general problem formulation of decentralized wireless resource allocation problems and address the problem by parameterizing the policies with statistical learning techniques. We further give some specific examples in this formulation. In Section \ref{sec:gnn}, we propose the formulation of Agg-GNN method to parameterize the policies. We investigate the important property of permutation equivariance of Agg-GNN with respect to the input graph structure in Section \ref{sec:permute}. Section \ref{sec:primal} gives the model-free primal-dual training method of Agg-GNN. Section \ref{sec:sim} shows results from numerical simulations that demonstrate the performance of our proposed Agg-GNN can outperform existing state-of-the-art strategies, as well as the verification of transference. 

%%%%%%%%%%%%%%%%%%%%%%%%%%%%%%%%%%%%%%%%%%%%%%%%%%%%%%%%%%%%%%%%%%%%%
%%%%%%%%%%%%%%%%%%%%%%%%%%%%%%%%%%%%%%%%%%%%%%%%%%%%%%%%%%%%%%%%%%%%
%%%   S  E  C  T  I  O  N   %%%%%%%%%%%%%%%%%%%%%%%%%%%%%%%%%%%%%%
%%%%%%%%%%%%%%%%%%%%%%%%%%%%%%%%%%%%%%%%%%%%%%%%%%%%%%%%%%%%%%%%%%%%
%%%%%%%%%%%%%%%%%%%%%%%%%%%%%%%%%%%%%%%%%%%%%%%%%%%%%%%%%%%%%%%%%%%%%
\section{Problem Formulation}
\label{sec:prob}
We consider a {cooperative} wireless system containing $m$ transmitters and $n$ receivers. Each transmitter $i\in\{1,2,\hdots,m\}$ is paired with a single receiver $r(i)\in\{1,2,\hdots,n\}$. {Multiple transmitters may be paired with the same receiver---e.g. a cellular uplink---or can be individually paired with a unique receiver---e.g. an ad-hoc or device-to-device network.} {Note that these are two generic network structures that can be used to define other specific network scenarios.} The channel state at discrete time instance $t\in\mbZ$ is characterized by a matrix $\bbH(t)\in\reals_+^{m\times m}$ whose element $|h_{ij}(t)| := [\bbH(t)]_{ij}$ stands for the channel condition between transmitter $i$ and receiver $r(j)$. {In addition to the channel state between a transmitter and receiver, we additionally consider} transmitter, or node, states represented by $\bbx(t)\in\reals^m$, {where $x_i(t) := [\bbx(t)]_i$ is the application state of the $i$-th node at time $t$}{, e.g. the  current  arrival  rate  of  traffic,  queue  length, and the  state  of  a  control  system in operation at the device.} {We consider a fast fading environment, in which both the channel state $\bbH(t)$ and node state $\bbx(t)$ randomly vary over $t$.} We use $m(\bbH,\bbx)$ to represent the probability distribution of the joint stochastic process $\{\bbH(t),\bbx(t)\}_{t\in\mbZ}$. This distribution is assumed stationary. The fading state $\bbH(t)$ may reflect, e.g., shadowing phenomena following a log-normal distribution, while the node state $\bbx(t)$ may reflect, e.g., packet arrivals following a Poisson distribution.

\begin{figure}
\centering
\usetikzlibrary{decorations.pathreplacing,angles,quotes,calc}

\begin{tikzpicture}
\draw (0 ,0) node[anchor=north] {t=0} -- (0,0.1);
\foreach \x in {1,2,...,6} { 
  
  \draw (\x ,0) node[anchor=north] {\x0} -- (\x,0.1);
  %\foreach \y in {.2,.4,.6,.8}{
 % 	\draw (\value{\x}+\value{y},0) node[anchor=north] {} -- (\value{\x}+\value{y},0.1);
   %};
};
\foreach \x in {0.1,0.2,..., 6.9} { 
  
  \draw (\x ,0) node[anchor=north] {} -- (\x,0.1);
  %\foreach \y in {.2,.4,.6,.8}{
 % 	\draw (\value{\x}+\value{y},0) node[anchor=north] {} -- (\value{\x}+\value{y},0.1);
   %};
};
\draw (7 ,0) node[anchor=north, yshift=-.1cm] {...} -- (7, 0.1);

\draw[->] (-0.5,0) -- (7.5,0);

%\fill[blue!50] (0, 0) rectangle ( 2, 0.1 );
\foreach \y in {0.04, 0.09, 0.14}{
\foreach \x in {0, 0.1, 0.4, 0.8, 1.3, 2 , 2.4, 2.8, 2.9, 3.4, 3.5, 3.9, 4, 4.1, 4.6, 5.2,5.3, 6.1, 6.4}{
\draw[color = blue!40, ultra thick] (\x,\y) -- (\x+0.1 ,\y);}
}

\foreach \y in {0.2, 0.25, 0.3}{
\foreach \x in{ 0, 1.4, 1.5, 2.8, 3.9, 4, 5.5, 5.6, 6.4 }{
\draw[color = red!40, ultra thick] (\x,\y) -- (\x+0.1,\y);}}

\foreach \y in {0.36, 0.41, 0.46}{
\foreach \x in { 0.2 ,0.3,  1.1, 1.2,  2.3, 2.4,  3.2, 3.3 , 3.4 , 4.3, 4.4, 5.1 ,5.2, 6.5, 6.6 }{
\draw[color = green!40, ultra thick] (\x ,\y) -- (\x + 0.1,\y);
}}

\draw [decorate,decoration={brace,amplitude=10pt}]
(0.2,0.48) -- (1.2,0.48) node [black,midway,xshift=-0,yshift=.3cm,above] 
{\footnotesize $\mathcal{T}_1$};

\end{tikzpicture}
\caption{Illustration of multi-level time scale and asynchronous operating times. The x-axis ticks reflect time instances in which the channel state changes. Colored markers reflect the time periods in which three asynchronous nodes update resource levels and send state information to neighbors. Working clocks may differ in both periodicity and offset.}
\label{fig_timescale}
\end{figure}
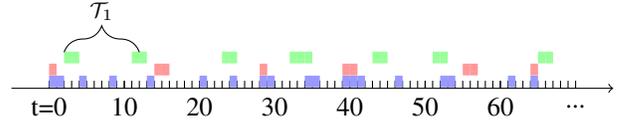

In a decentralized network setting, frequent communication overhead is required to keep all the network nodes operating under a synchronous clock without a central controller. We design resource allocation policies for the more general asynchronous scenarios by modeling heterogeneous working patterns for each node. The working status of each node varies relative to the more granular reference time index $t \in\mbZ$. More specifically, we denote the set of active nodes at time $t$ as $\ccalA(t)\in\{1,2,\hdots, m\}$. This indicates that only nodes $i\in\ccalA(t)$ can take actions, such as sending information to neighboring nodes, and make decisions, such as updating its resource allocation strategy. In Figure \ref{fig_timescale} we show an example of this asynchronous structure. The time $\ccalT_i$ between active time indexes for a node $i$ reflects the time scale at which it makes resource allocation decisions.

Technically, the matrix $\bbH(t)$ can be dense as there exists a measurable channel state between all active wireless devices. However in practical large network setting, unpaired transmitters and receivers are more likely to be too far from each other to cause interference, which results in negligible channel quality between them. {We thus consider sparse forms of $\bbH(t)$ by imposing a threshold value $\eta_0$. Therefore, for a specific node $i$ we focus on the set of active nodes that may cause non-negligible interference to its signal transmission and subsequently define the neighborhood of node $i$ as $\ccalN_{i}(t)=\{ [\bbH(t)]_{ij} \geq \eta_0, j\in \ccalA(t)\}$.} Furthermore, we define a sparsifying matrix as $\bbQ(t)\in\{0,1\}^{m\times m}$, with $[\bbQ(t)]_{ij}=1$ if $j\in\ccalN_i(t)$ and 0 otherwise. The sparsity of $\bbQ(t)$ may reflect the channel measurement bandwidth of a device. We then consider a limited channel state matrix $\tbH(t)$ defined as the element-wise product
\begin{align}
\label{eqn:limitchannel}
    \tbH(t) := \bbH(t) \circ \bbQ(t).
\end{align}
We note that this $\tbH(t)$ incorporates both current channel state as well as the asynchronous working patterns of neighboring nodes. Therefore it can represent actual communicating devices at each time slot.

%\blue{it's receiver $r(i)$'s neighborhood}
%\blue{define a neighborhood $\ccalN_i(t)$ (i.e. $\bbH$ is technically dense but in practice is sparse in large networks)}

A receiver $r(i)$ can also receive signals from larger neighborhoods with some delay---e.g. the signal of its neighbors' neighbors can be received after one additional time step. Recursively, we can then define node $i$'s $k$-hop neighborhood as $\ccalN_i^k(t) := \{j'\in\ccalN_j(t-k+1), j\in\ccalN_i^{k-1}(t) \cap \ccalA(t) \}$. Based on this notation, we define locally available observations at node $i$ to include states that can either be observed directly or obtained through delayed information exchanges with neighboring nodes $j\in \ccalN_{i}(t)$. Locally available history information collected at node $i$ at time $t$ therefore can be defined as
\begin{align} \label{eqn:local_info}
\ccalH_{i}(t):=\bigcup_{k=0}^{K-1} \Big\{ [\bbH(t-k+1)]_{jj'},& [\bbx(t-k)]_{j'} \mid  \\
&  j\in\ccalN_{i}^{k-1}(t), j'\in\ccalN_{i}^{k }(t) \Big\}. \nonumber
% \ccalH_{i}(t):=\bigcup_{k=0}^{\lceil\frac{K}{n_{ex}} \rceil-1} \bigcup_{n=1}^{n_{ex}} \Big\{ [\bbH(t-k)]_{jj'},& [\bbx(t-k)]_{j'} \mid  \\
% &  j\in\ccalN_{i}^{n-1}(t), j'\in\ccalN_{i}^{n}(t) \Big\}. \nonumber
\end{align}
We limit the complexity of information exchanges in \eqref{eqn:local_info} by setting the maximal neighborhood range as $K$ and channel threshold $\eta_0$, which directly influence the amount of information contained in $\ccalH_i(t)$.

Our goal is to {determine a local resource allocation policy  {$\bbp_i(\ccalH_i(t))$}} as a mapping of a node's local history information to an instantaneous resource allocation %, i.e. $\bbp_i(t)=\bbp(\ccalH_{i}(t))$  
under some specific constraints for each node $i$. When each node is allocated with resource under this mapping at time $t$, we can denote all the decisions together as  {$\bbP(\ccalH(t))=[\bbp_1(\ccalH_{1}(t)), \bbp_2(\ccalH_{2}(t)),\hdots,\bbp_m(\ccalH_{m}(t))]$}. %$\bbP(\ccalH(t))=[\bbp(\ccalH_{1}(t)), \bbp(\ccalH_{2}(t)),\hdots,\bbp(\ccalH_{m}(t))]$. 
Together with system state pair $(\bbH(t), \bbx(t))$, a collection of instantaneous performance feedbacks $\bbf\left(\bbP(\ccalH(t)), \bbH(t),\bbx(t)\right)$ are observed. Fast variations in channel and node states indicates that we need to design with respect to a long term average performance, which can be evaluated as an expectation {with respect to the channel and node states}. %\red{Explain that the observation time period is longer compared with the coherence time.} 
Therefore, we {define} $\bbr \in \reals^{m}$ {as the expected reward under the decision set $\bbP(\ccalH)$} over all random states, i.e.,
\begin{align}
    \bbr&:=\mathbb{E} \left[ \bbf\left(\bbP(\ccalH) , \bbH,\bbx\right) \right] %\nonumber \\
    &= \int \bbf\left(\bbP(\ccalH), \bbH,\bbx\right) \text{d}m(\bbH,\bbx) . 
    \label{eqn:average_reward}
\end{align}
Here the expectation is taken over both the current random states $\bbH, \bbx$ and the history state information $\ccalH$ as determined by the probability distribution $m(\bbH,\bbx)$ of the stochastic process $\{\bbH(t),\bbx(t)\}_{t\in\mbZ}$. Observe that since the process is stationary, the expectation in \eqref{eqn:average_reward} is independent of the time index $t$ and for that reason we have dropped the time index.

The optimal policy is the one that maximizes a global utility $u_0: \reals^m \rightarrow \reals$ while satisfying a set of system constraints $\bbu: \reals^m \rightarrow \reals^q$ with respect to the long term reward $\bbr$ in \eqref{eqn:average_reward}.  With all the notations above, the optimal resource allocation policy $\bbP^*(\ccalH)$ and associated average rewards $\bbr^*$ are given by:
\begin{equation}\label{eqn:prob}
\begin{aligned}
    &[\bbP^*(\ccalH), \bbr^*] := &\argmax_{\bbP(\ccalH), \bbr}   \quad &u_0(\bbr)   \\
    &&\st \quad& \bbr =\mathbb{E}\left[\bbf\left( \bbP(\ccalH), \bbH,\bbx\right)\right], \\
   &&&  \bbu(\bbr)\geq\mathbf{0},\quad \bbP(\ccalH)\in\mathcal{P}.
\end{aligned}
\end{equation}

The joint {reward} function $\bbf$ is often non-convex, which makes the solution of \eqref{eqn:prob} intractable. Heuristic methods are often used to {find local stationary points of \eqref{eqn:prob}}, e.g. \cite{shi2011iteratively}, {but necessarily require explicit model knowledge}. In our proposed policy, we implement a data-driven method to update the resource allocation strategy based on the observations from previous policy. With no prior knowledge of the model, we can consider this as a constrained statistical learning problem. The resource allocation function for each node $\bbp_i(\ccalH_i)$ can be {substituted with} a common function $\bm{\phi}(\ccalH_{i}, \bbA)$, where $\bm{\phi}$ is a vector-valued function family with input a shared parameter $\bbA \in \reals^s$. With $\bm{\Phi}(\ccalH, \bbA)= [\bm{\phi}(\ccalH_1,\bbA), \bm{\phi}(\ccalH_2,\bbA), \hdots, \bm{\phi}(\ccalH_m,\bbA)]$, the optimization problem \eqref{eqn:prob} therefore can be reformulated as
\begin{equation}\label{eqn:opt}
\begin{aligned}
    &[\bbA^*, \bbr^*] := &\argmax_{\bbA, \bbr}   \quad &u_0(\bbr)   \\
    &&\st \quad& \bbr =\mathbb{E}\left[\bbf\left( \bm{\Phi}(\ccalH, \bbA), \bbH,\bbx\right)\right], \\
   &&&  \bbu(\bbr)\geq\mathbf{0}. %,\quad \bbP(\ccalH)\in\mathcal{P}.
\end{aligned}
\end{equation}

The benefits of reformulating the problem as \eqref{eqn:opt} is that the solution set is now a variable $\bbA$ with controllable dimension $s$, instead of dense sets containing $\bbH$ and $\bbx$. Furthermore, the function is shared across all the nodes and therefore requires the optimization of only a single parameter $\bbA$. This avoids the need for learning a separate policy for each node and improves the optimization efficiency as well as permitting a transferability of the designed policy to new network nodes. The constraint $\bbP(\ccalH)\in\ccalP$ has been removed for this can be easily achieved by a projection operation on $\bm\phi(\ccalH_i,\bbA)$. {It can be shown, in fact, that selecting an appropriate function family, such as Fully Connected Neural Networks (FCNNs), with a large enough parameter dimension $s$ can make the problems in \eqref{eqn:prob} and \eqref{eqn:opt} almost the same} \cite{de2018team}. However, if we choose to employ a FCNN, the dimension $s$ needed would grow linearly with the wireless network size. {Moreover, traditional neural networks are not suited for parameterizing a policy $\bbp_i(\ccalH_i)$ for each node whose input $\ccalH_i$ is of varying dimension---that is, the amount of neighborhood information collected at each node varies across nodes and network topologies.} 

{In this paper, we} propose to use  Aggregation Graph Neural Networks (Agg-GNNs) over random edge graphs which can scale well and find good solutions to \eqref{eqn:opt} in larger networks with limited parameters. {The constrained resource allocation problem considered in this paper may capture any type of resource, such as power, time slots or frequency bands. Each of these resource unit structures can be encoded in the neural network architecture and therefore can be allocated with our method.} We proceed to present some examples of practical resource allocation problems which can be written in the form of \eqref{eqn:prob} before we introduce the formulation of graph neural networks in the following section.

\begin{remark}
\label{remark_coherence}
\normalfont
It is customary to assume that the state variables $\bbH(t), \bbx(t)$ are independent and identically distributed (i.i.d.). We remove that assumption here and allow for correlated channel models. In the numerical experiments of Section \ref{sec:sim} we adopt a Rayleigh fading model in which $h_{ij}(t)$ is a complex normal with zero mean, variance $\sigma^2$, and uncorrelated real and imaginary parts. The channel's coherence is described by an innovation factor $\delta$. Formally, let $w_{ij}(t)$ be an i.i.d. sequence of complex normal random variables with mean $\mbE[w_{ij}(t)]=0$ and with real and imaginary parts that satisfy $\mbE[\Re{w_{ij}(t)}^2]=\mbE[\Im{w_{ij}(t)}^2]=\sigma^2$ and $\mbE[\Re{w_{ij}(t)}\Im{w_{ij}(t)}]=0$. Then, the channel $h_{ij}(t)$ evolves according to 
\begin{align}
\label{eqn:Gau-Mar}
   h_{ij}(t+1) 
      ~=~ \sqrt{1-\delta}\, h_{ij}(t)
             ~+~ \sqrt{\delta}\, w(t+1) ~.
\end{align}
The innovation factor $\delta$ controls the rate of change of the channel absolute values $|h_{ij}(t)|$ that form the entries of the matrix $\bbH(t)$. When $\delta=1$ channel realizations are i.i.d., and when $\delta=0$ channel realizations are constant. Since our goal is to design policies that take advantage of global information (see Section \ref{sec:gnn}), this will play a role in observed performance. We expect smaller $\delta$ to lead to problems where our learned policies perform better. Our numerical experiments corroborate that this is true (see Section \ref{sec_nn1}.).
\end{remark}

\subsection{Examples}
\label{sec:eg}
\subsubsection{Dynamic Power Allocation in AWGN Channels}
\label{sec:dynpowAWGN}
Transmitters communicate with associated receivers over a shared AWGN channel. The common objective is to maximize the sum of the channel capacity of each receiver under noise and interference. Channel state $\bbH$ characterizes the channel condition of each possible link while the node state $\bbx$ takes no effect. The instantaneous sum rate can be written as
\begin{align}
    f_i\left( \bbP(\ccalH), \bbH,\bbx\right) =  \log\left(1+\frac{|h_{ii}|^2 p_i(\ccalH_{i})}{1+ \sum\limits_{j\in\ccalN_{i}}|h_{ij}|^2 p_j(\ccalH_{j})}\right)
\end{align}
% The Signal-to-Interference-Ratio at receiver $i$ at time $t$ is given by:
% \begin{align}
%     \gamma_{it}(\bbp(\ccalH_{it}), \ccalH_{it})=\frac{h_{ii}(t)[\bbp(\ccalH_{it})]_i}{1+\sum\limits_{j\in\ccalN_{it}}h_{ij}(t) [\bbp(\ccalH_{it})]_j }.
% \end{align}
% The channel capacity therefore is
% \begin{align}https://www.overleaf.com/project/5e377f2e6595130001dd9774
%     f_i(\bbp(\ccalH_{it});\ccalH_{it})=\log(1+\gamma_{it}(\bbp(\ccalH_{it}), \ccalH_{it})).
% \end{align}
The utility function can be set as $u_0(\bbr)=\bbr^T\bm{1}$ or $u_0(\bbr)=\sum_{i=1}^m \log(r_i)$ if fairness is considered. The constraint function can be used to set a lower bound $\bbc_{min}$ for the sum capacity, i.e. $\bbu(\bbr)=\bbr- \bbc_{min}\geq \bm{0}$.  

\subsubsection{Dynamic Power Allocation with User Demands}
When considering the data arrival rate of each node, $\bbx$ can be incorporated to form the utility and constraint functions as
\begin{align}
    f_i\left(  \bbP(\ccalH), \bbH,\bbx\right) = \log\left(1+\frac{|h_{ii}|^2 p_i(\ccalH_{i})}{1+ \sum\limits_{j\in\ccalN_{i}}|h_{ij}|^2 p_j(\ccalH_{j})}\right)-x_i.
\end{align}
Therefore, the common utility function can be defined as $u_0(\bbr)=\bbr^T\bm{1} +\bbx^T \bm{1}$ while the constraint function is defined as $\bbu(\bbr)=\bbr^T\bm{1}\geq 0$.

\subsubsection{Distributed Random Access}
%\blue{we can make this a general distributed random access}
In wireless local area networks and cellular systems, transmitters contend to get the access to a common access point (AP). The is commonly done in a distributed manner via random access. Consider a function $q$ with binary output that determines channel access decision based on transmission powers and channel states. The actual transmission rate for transmitter $i$ depends upon whether or not collisions have occurred and can be written as
\begin{align}
&\nonumber f_i\left(  \bbP(\ccalH), \bbH \right)\\ 
&=c_i(\bbh_i ,p_i(\ccalH_{i})) q(p_i(\ccalH_{i}), \bbh_i)  \prod_{j\neq i}(1-q(p_j(\ccalH_{j}), \bbh_j)),
\end{align}
where the function $c_i(\bbh_i, p_i(\ccalH_{i}))$ defines the transmission rate, the exact form of which is determined by the physical layer.
The utility function $u_0(\bbr)=\bbr^T\bm{1}$ is set as the sum of the actual transmission rates. The constraint function can be set as $\bbu(\bbr)= \bbr^T \bm{1}\geq 0$.

%{When considering the resource allocation in a wireless control system, transmitters send plant state information to a receiver to get control feedback over a common random access channel. $q$ function gives a probability of collision between two transmitters, with transmission powers and channel states as inputs. $\bbx$ here stands for the plant state which either evolves with stable gain $0<\gamma_c<1$ or unstable gain $\gamma_o>1$. The cost function can be written as
%\begin{align}
%    &f_i\left(  \bbP(\ccalH), \bbH,\bbx\right)\\ &=-\sum\limits_{i=1}^m\left(\prod_{j\neq i}(1-q(p_i(\ccalH_{i}), h_{ii}, p_j(\ccalH_{j}), h_{ij}))\right)\\
%    & \qquad \qquad (\gamma_c^2-\gamma_o^2)x_i^2 +\gamma_o^2 x_i^2.
%\end{align}
%$u_0(\bbr)=-\bbr^T\bm{1}$ is used to define the common utility function. Constraint function can be set as the maximum cost threshold $\bbc_{max}$, i.e. $\bbu(\bbr)=\bbc_{max}-\bbr\geq0$.}

%%%%%%%%%%%%%%%%%%%%%%%%%%%%%%%%%%%%%%%%%%%%%%%%%%%%%%%%%%%%%%%%%%%%%
%%%%%%%%%%%%%%%%%%%%%%%%%%%%%%%%%%%%%%%%%%%%%%%%%%%%%%%%%%%%%%%%%%%%
%%%   S  E  C  T  I  O  N   %%%%%%%%%%%%%%%%%%%%%%%%%%%%%%%%%%%%%%
%%%%%%%%%%%%%%%%%%%%%%%%%%%%%%%%%%%%%%%%%%%%%%%%%%%%%%%%%%%%%%%%%%%%
%%%%%%%%%%%%%%%%%%%%%%%%%%%%%%%%%%%%%%%%%%%%%%%%%%%%%%%%%%%%%%%%%%%%%
\section{Aggregation Graph Neural Networks}
\label{sec:gnn}
To paramaterize the power allocation policy $\bbP(\ccalH)$ in a manner that supports distributed decision making, we consider the use of a localized deep learning architecture called the Aggregation Graph Neural Network (Agg-GNN). We begin by re-contextualizing the wireless state variables $\bbH$ and $\bbx$ as states on a collection of random time-varying graphs. % based on the GNN architecture. 
Consider a graph $\ccalG := \{\ccalV, \ccalE\}$ with nodes $\ccalV:=\{1,2,\hdots,m\}$ corresponding to the $m$ transmitters in the wireless network and edges $\ccalE := \{(i,j) \mid i,j=1,2,\hdots,m\}$, where the weighted edge $(i,j)$ corresponds to the strength of the channel between a transmitter $i$ and receiver $r(j)$. 
% The resulting graph structure is illustrated more clearly in Figure \ref{fig_g_signal}. 
From here, we may reinterpret $\bbx(t)=[x_1; x_2;\hdots; x_m](t)$ as a signal supported on the nodes and the sparse channel matrix $\tbH(t)$ as the weighted {adjacency matrix} of a graph $\ccalG(t)$. We note that this graph structure considers both current channel condition and the asynchronous activation patterns of each node. In graph signal processing literature, the matrix $\tbH(t)$ is called a graph shift operator (GSO) \cite{sandryhaila2014big}. 

\subsection{Wireless Graph Aggregation Sequence}
% Similar to the above concept interpretation, the process of signal transmission over the wireless channel at time $t$ can be seen as an application of the GSO to the graph signal in the above context. Consider a set of delayed states $\{x_j(t-1)\}_{j=1}^m$ transmitted over an additive wireless channel by transmitters $j=1,\hdots,m$, respectively. The signal received at receiver $r(i)$ at time $t$ is impacted by fading as
% \begin{equation}\label{eq_transmit_i}
%   y^{(1)}_i(t) = \sum_{j \in \ccalN_i(t)} h_{ij}(t) x_j(t-1).
% \end{equation}

The basis of the Agg-GNN parameterization comes from a graph diffusion, or aggregation, operation. A graph aggregation of the node states $\bbx(t)$ can be represented with an application of the GSO matrix $\bbH(t)$, i.e.,
\begin{equation}\label{eq_transmit_i}
   \bby^{(1)}(t)  := \tbH(t) \bbx(t-1).
\end{equation}
Observe that the $i$-th element of $\bby^{(1)}(t)$ in \eqref{eq_transmit_i} can be obtained locally at node $i$ by a weighted aggregation of node state information from its immediate neighbors, i.e. $y^{(1)}_i(t) = \sum_{j \in \ccalN_i(t)}| h_{ij}(t)| x_j(t-1)$. The aggregated signal $\bby^{(1)}(t)$ can be further aggregated at next transmission step $t+1$, corresponding to another graph shift operation $\tilde\bbH(t+1)$. The resulting aggregated signal is given by $\bby^{(2)}(t+1) = \tilde\bbH(t+1) \bby^{(1)}(t) = \tilde\bbH(t+1) \tilde\bbH(t) \bbx(t-1)$. %Again, each element can be obtained locally at each node through two rounds of exchanges with active neighbors. 
After $K$ successive transmissions, we can obtain a sequence of shifted graph signals $\bby^{(0)}(t), \bby^{(1)}(t), \hdots, \bby^{(K-1)}(t)$, where $\bby^{(0)}(t) := \bbx(t)$ and the following elements defined as
\begin{align}
\label{eq:aggregation1}
   \bby^{(k)}(t)&:= \tbH(t)\bby^{(k-1)}(t-1)=\left[ \prod_{t'=0}^{k-1} \tilde\bbH(t-t') \right]\bbx(t-k).
\end{align}
We again emphasize that the $i$-th element of any aggregate $\bby^{(k)}(t)$ can be obtained locally by node $i$ over $k$ time steps solely through local exchanges with its immediate neighbors $j \in \ccalN_i(t')$ for $t' \in \{t-k+1,\hdots,t\}$. Thus we consider the local aggregation sequence held at node $i$ as
\begin{align}
\label{eqn:agg}
     \bby_{i}(t)=[y^{(0)}_i(t); y^{(1)}_i(t); \hdots; y^{(K-1)}_i(t)].
\end{align}
We call $\bby_{i}(t)$ in \eqref{eqn:agg} the \emph{wireless aggregation sequence} collected by node $i$, as the elements successively collect information of the active node and channel state information of the global wireless network. Further observe that $\bby_{i}(t)$ is constructed using precisely the local delayed information structure $\ccalH_i(t)$ defined in \eqref{eqn:local_info} and thus can be used as an input to a local allocation policy $\bbphi(\ccalH_i, \bbA)$.  This aggregated signal sequence also reflects the topology information because the components of the time sequence depend on the graph structure. The $k$-th element corresponds to the aggregated information from $k$-hop neighbors.  We further notice that each active node transmits its aggregated information once per time slot as an overhead message. While inactive nodes cannot transmit signals, they can still receive aggregated information which can be forwarded to their neighboring nodes during their next active phase.

%%%%%%%%%%%%%%%%%%%%%%%%%%%%%%%%%%%%%%%%%%%%%%%%%%%%%%%%%%%%%%%%%%%
%%   F   I   G   U   R   E   %%%%%%%%%%%%%%%%%%%%%%%%%%%%%%%%%%%%%%
%%%%%%%%%%%%%%%%%%%%%%%%%%%%%%%%%%%%%%%%%%%%%%%%%%%%%%%%%%%%%%%%%%%
\begin{figure*}
\centering
% !TEX root = ../root.tex

\tikzstyle{empty node} = [ circle, 
                           draw = black,
                           text = black, 
                           minimum size = 0.65*\unit]
                           
\tikzstyle{empty node small} = [ circle, 
                           draw = black,
                           text = black, 
                           minimum size = 0.5*\unit]

\tikzstyle{blue node} = [ empty node, 
                         fill = blue!50,
                         draw = blue!50,
                         text = white]

\tikzstyle{red node} = [ empty node, 
                         fill = red!50,
                         draw = red!50,
                         text = white]

\tikzstyle{gray node} = [ empty node, 
                         fill = black!50,
                         draw = black!50,
                         text = white]
                         
\tikzstyle{blue node small} = [ empty node small, 
                         fill = blue!50,
                         draw = blue!50,
                         text = white]
                         
\tikzstyle{gray node small} = [ empty node small, 
                         fill = black!50,
                         draw = black!50,
                         text = white]

\tikzstyle{blue node small next} = [ empty node small, 
                         fill = blue!40,
                         draw = blue!40,
                         text = white]

\tikzstyle{red node small} = [ empty node small, 
                         fill = red!50,
                         draw = red!50,
                         text = white]

\tikzstyle{gray node small next} = [ empty node small, 
                         fill = black!40,
                         draw = black!40,
                         text = white]

\tikzstyle{edge} = [shorten >=0pt, shorten <=0pt]

\def \myfactor {1.00}
\def \unit     {\myfactor cm}

\def \radius   {2.5}

{\fontsize{7}{7}\selectfont\begin{tikzpicture}[scale = \myfactor]

 % Nodes in the left pentagon 
  \node                             [blue node small]  at (-20,0) (ic) {i}; % Defines center of pentagon
  \path (ic) ++ (  30:\radius/2) node [blue node small] (1c)   {$j_1$}; % ++ ( 0.5,  0.5) node {$x_1$}; %;
  \path (ic) ++ ( 140:\radius/2+0.5/2 ) node [blue node small] (2c)   {$j_2$}; %  ++ ( 0.6,  0.4) node {$x_2$}; %;
  \path (ic) ++ (250: \radius/2 -0.5/2) node [blue node small] (3c)   {$j_3$}; %  ++ (-0.6,  0.4) node {$x_3$}; %;
  \path (ic) ++ (315:\radius/2) node [gray node small] (4c)   {$j_4$}; %  ++ (-0.5,  0.5) node {$x_4$}; %;
  
  \path (3c) ++ ( 210:\radius ) node [gray node small next] (k1c)   {$k_1$};
  
  \path (3c) ++ (  320 :\radius+0.5 ) node [blue node small next] (k2c)   {$k_2$};

%   % Edges connecting left pentagon  
\path[-stealth] (1c)  edge [edge] node {} (ic);
  \path[-stealth] (2c)  edge [edge] node {}  (ic);
  \path[-stealth] (3c)  edge [edge] node {}  (ic);
  \draw [shorten >=0pt,dashed] (4c) --node{} (ic);
  
\draw [shorten >=0pt,dashed] (k1c) --node{} (3c);  
\path[-stealth] (k2c)  edge [edge] node []{}  (3c);

  % Nodes in the left pentagon 
  \node                             [red node small] at (-14,-1) (ia) {i}; % Defines center of pentagon
  \path (ia) ++ (  30:\radius) node [blue node small] (1a)   {$j_1$}; % ++ ( 0.5,  0.5) node {$x_1$}; %;
  \path (ia) ++ ( 140:\radius+0.5) node [blue node small] (2a)   {$j_2$}; %  ++ ( 0.6,  0.4) node {$x_2$}; %;
  \path (ia) ++ (250:\radius-0.5) node [blue node small] (3a)   {$j_3$}; %  ++ (-0.6,  0.4) node {$x_3$}; %;
  \path (ia) ++ (315:\radius) node [gray node small] (4a)   {$j_4$}; %  ++ (-0.5,  0.5) node {$x_4$}; %;

  % Edges connecting left pentagon  
  \path[-stealth] (1a)  edge [edge] node [fill=white,inner sep=1pt]{$h_{ij_1}(t)x_{j_1}(t-1)$} (ia);
  \path[-stealth] (2a)  edge [edge] node [fill=white,inner sep=1pt]{$h_{ij_2}(t)x_{j_2}(t-1)$}  (ia);
  \path[-stealth] (3a)  edge [edge] node [fill=white,inner sep=1pt]{$h_{ij_3}(t)x_{j_3}(t-1)$}  (ia);
  \draw [shorten >=0pt,dashed] (4a) --node{} (ia);

  % Nodes in the left pentagon 
  \node                             [red node small]   at (-7,0)(ib)  {i}; % Defines center of pentagon
  \path (ib) ++ (  30:\radius/2) node [blue node small] (1b)   {$j_1$}; % ++ ( 0.5,  0.5) node {$x_1$}; %;
  \path (ib) ++ ( 140:\radius/2+0.5/2 ) node [blue node small] (2b)   {$j_2$}; %  ++ ( 0.6,  0.4) node {$x_2$}; %;
  \path (ib) ++ (250: \radius/2 -0.5/2) node [blue node small] (3b)   {$j_3$}; %  ++ (-0.6,  0.4) node {$x_3$}; %;
  \path (ib) ++ (315:\radius/2) node [gray node small] (4b)   {$j_4$}; %  ++ (-0.5,  0.5) node {$x_4$}; %;
  
  \path (3b) ++ ( 210:\radius ) node [gray node small next] (k1b)   {$k_1$};
  
  \path (3b) ++ (  320 :\radius+0.5 ) node [blue node small next] (k2b)   {$k_2$};

%   % Edges connecting left pentagon  
\path[-stealth] (1b)  edge [edge] node {} (ib);
  \path[-stealth] (2b)  edge [edge] node {}  (ib);
  \path[-stealth] (3b)  edge [edge] node {}  (ib);
  \draw [shorten >=0pt,dashed] (4b) --node{} (ib);
  
  \path[-stealth] (k1b)  edge [edge] node [fill=white,inner sep=1pt]{$h_{j_3k_1}(t-1)x_{k_1}(t-2)$} (3b);
  \path[-stealth] (k2b)  edge [edge] node [fill=white,inner sep=1pt]{$h_{j_3 k_2}(t-1)x_{k_2}(t-2)$}  (3b);

\end{tikzpicture}} 
\caption{
An example network at time $t$ is shown with gray nodes inactive and blue nodes active (left). Node $i$ collects information from 1-hop neighbors (middle) and 2-hop neighbors (right). At time $t$, the neighbors of node $i$, which are close enough to $i$ and also active, send directly their delayed node information $\{x_{j_1}(t-1), x_{j_2}(t-1),x_{j_3}(t-1)\}$ to node $i$ which form $y_i^{(1)}(t)$. The signal $y_{j_3}^{(1)}(t-1)$ sent from node $j_3$ to node $i$ is actually formed by delayed information $\{x_{k_1}(t-2), x_{k_2}(t-2)\}$  from node $i$'s 2-hop neighbors $k_1$ and $k_2$, which may be inactive at time $t$. }
\label{fig_g_signal}
\end{figure*}
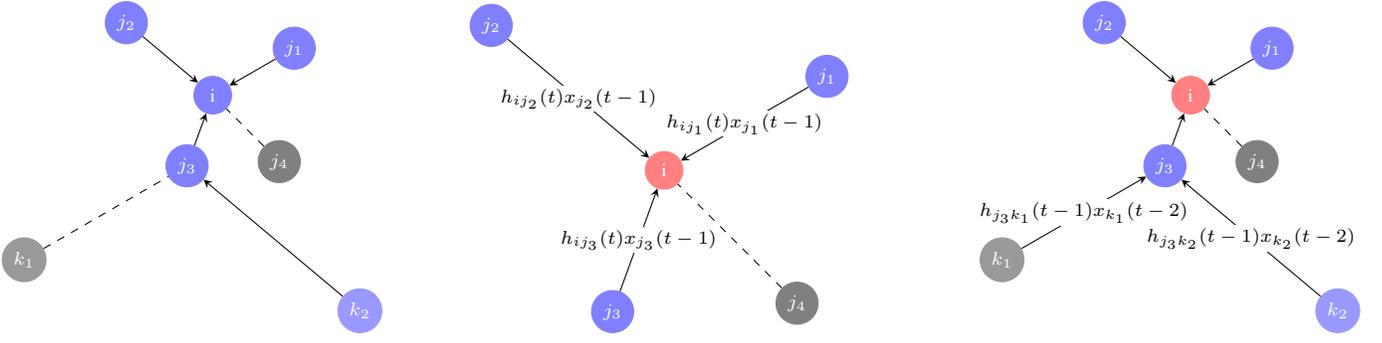

\subsection{Graph Neural Networks}
The resulting Agg-GNN resource allocation policy can be computed by node $i$ from its local aggregation sequence $\bby_i(t)$. Observe that, after $K-1$ successive aggregations over the wireless network, each node obtains a local temporally structured sequence of state information that captures the fading channel patterns of both its immediate neighborhood and delayed fading channel patterns of the global network.
Given the temporal structure of $\bby_i(t)$, we implement a standard Convolutional Neural Network (CNN) architecture with $L$ layers. Formally, the architecture begins with a linear transformation to produce an intermediate output, which is followed by a pointwise nonlinear function. By applying this procedure recursively, for node $i$ at the $l$-th layer we can get
\begin{align}
    \bby_{il} =\sigma_l[\bbv_{il}]=\sigma_l[\bbA_l  \bby_{i(l-1)} ].
\end{align}
%for the ease and generality of presentation, we temporally omit the node index here.
Consider multiple features per layer, the output of the $l-1$-th layer can be written as {a combination of feature sequences: } $\bby_{i(l-1)}:=[ \bby_{i(l-1)}^1;\hdots;\bby_{i(l-1)}^{F_{l-1}} ] $, where $F_{l-1}$ represents the number of features in the $l-1$-th layer. Likewise, the intermediate output of the $l$-th layer can be written as $\bbv_{il}:=[ \bbv_{il}^1;\hdots;\bbv_{il}^{F_l} ]$. Let $\bm\alpha_l^{fg}:=[[\bm\alpha_l^{fg}]_1;\hdots;[\bm\alpha_l^{fg}]_{K_l}]$ be the coefficients of a $K_l$-tap linear filter which is used to process the $g$-th feature of $l-1$-th layer to produce feature $\bbv_{il}^{fg}$ with convolution. This can be explicitly written as
\begin{align}\label{eq_gnn1}
[\bbv_{il}^{fg}]_n=[\bm\alpha_l^{fg} * \bby_{i(l-1)}^g]_n =\sum\limits_{k=1}^{K_l}[\bm\alpha_l^{fg}]_k [\bby_{i(l-1)}^g]_{n-k}.
\end{align}
The $l$-th layer therefore produces $F_{l-1}\times F_l$ features $\bbv_{il}^{fg}$ with the same size as the input. By aggregating all features and passing through a pointwise nonlinearity function $\sigma$, the $l$-th layer final output is 
\begin{align}
\bby_{il}=\sigma_l[\bbv_{il}^f] =\sigma_l\left[\sum\limits_{g=1}^{F_{l-1}}\bbv_{il}^{fg}\right]=\sigma_l\left[\sum\limits_{g=1}^{F_{l-1}} \bm\alpha_l^{fg} * \bby_{i(l-1)}^g\right].
\end{align}
At the first layer, we can rewrite the output features from input on node $i$ to show the involvement of graph structure as
\begin{align}
 \nonumber&[\mathbf{y}_{i1}^{fg}]_n = \\
 &\sigma_1\left[\sum_{k=1}^{K_1} [\bm\alpha_1^{fg}]_k \left[\prod\limits_{m=0}\limits^{n-k-1} \tilde\bbH(t-m ) \bbx(t-n+k)  \right]_i \right].
\end{align}
The filter parameters are shared across all the nodes. Grouping all the parameters, we can get a filter tensor as $\bbA=\{\bm\alpha_l^{fg}\}_{l,f,g}$. Therefore, the operator can be defined as
\begin{equation}
\label{eqn:output}
    \bm{\phi}(\ccalH_{i}(t), \bbA)= \bby_{iL}(t).
\end{equation}
{The output can be seen as the decentralized resource allocation action at node $i$ at time $t$. Function $\bm\phi$ here is shared across all the nodes with the same filter tensor $\bbA$, which means we train a common GNN for all nodes other than a node-wise allocation function. Each node input its own local information sequence and get their corresponding resource allocation strategy. This generality also leads to the transferability of our trained GNN, which will be discussed in the following section.} The detailed operation of this resource allocation process for each node is given in Algorithm \ref{alg:Sequence}. {During the training phase, Algorithm 1 is implemented at each node to give performance feedbacks when updating parameters in Agg-GNN.}

%\blue{Comment that the parameters are the same for each node, i.e. we are defining a single GNN that is used locally by each transmitter to select power. Remark 3}

\begin{algorithm}[t]
\caption{{Resource Allocation at Node $i$}}
\label{alg:Sequence}
\begin{algorithmic}[1]
\FOR{$t\in \mbZ$ }
\IF{Node $i$ is active, $i \in \ccalA(t)$}
\STATE{Observe node state $x_i(t)$ and set $y^{(0)}_i(t) = x_i(t)$.}
\STATE{{Node $i$ transmits sequence $\{y^{(k)}_i(t-1)\}_{k=0}^{K-2}$ to transmitter $j \in \ccalN_i(t)$}.}
\STATE{Node $i$ forms aggregation sequence $\bby_i(t)$ based on information from its active neighbors.}
\STATE{Node $i$ updates resource level $p_i(t) = \phi(\ccalH_i(t),\bbA) = \bby_{iL}(t)$. }
\ELSE
\STATE{Node $i$ receives information from its active neighbors and forms aggregation sequence $\bby_i(t)$.}
\STATE{Node $i$ keeps resource level $p_i(t) = p_i(t-1)$.}
\ENDIF
\ENDFOR
\end{algorithmic}
\end{algorithm}

\begin{remark}\label{remark_aggregation}\normalfont
%\blue{Add remark about ``over-the-air'' computation of aggregation sequence/GNN vs. direct exchange of channel and node states directly between transmitters and the positives/negatives of both. Maybe this will end up as part of the main text instead of a remark, but lets write it as a remark for now.}
{
Observe that the formation of the aggregation sequence in Algorithm \ref{alg:Sequence} requires nodes exchange both channel state information $h_{ij}(t)$ and node state information $x_i(t)$ with neighboring devices a total of $K$ times. We emphasize that, while a total of $K$ exchanges are performed, only one exchange is needed per time instance $t$, thus rendering no additional overhead relative to decentralized methods with only single-hop neighborhood exchanges} \cite{shi2011iteratively}. Moreover, the number of exchanges can be controlled via the channel threshold $\eta_0$ that sets the sparsity of the graph.
\end{remark}

%%%%%%%%%%%%%%%%%%%%%%%%%%%%%%%%%%%%%%%%%%%%%%%%%%%%%%%%%%%%%%%%%%%%%
%%%%%%%%%%%%%%%%%%%%%%%%%%%%%%%%%%%%%%%%%%%%%%%%%%%%%%%%%%%%%%%%%%%%
%%%   S  E  C  T  I  O  N   %%%%%%%%%%%%%%%%%%%%%%%%%%%%%%%%%%%%%%
%%%%%%%%%%%%%%%%%%%%%%%%%%%%%%%%%%%%%%%%%%%%%%%%%%%%%%%%%%%%%%%%%%%%
%%%%%%%%%%%%%%%%%%%%%%%%%%%%%%%%%%%%%%%%%%%%%%%%%%%%%%%%%%%%%%%%%%%%%
\section{Permutation Equivariance}
\label{sec:permute}
The Agg-GNN architecture detailed in \eqref{eq_gnn1}-\eqref{eqn:output} is advantageous for decentralized resource allocation not only in its distributed inference capabilities, but perhaps just as critically in its adherence to an essential property of wireless networks. In particular, general GNN architectures are known to maintain an \emph{equivariance to permutations} of the underlying graph \cite{gama2019convolutional}. This property is indeed critical for autonomous decision making policies in wireless networks, which are inherently dynamic in their underlying topology. Given that learning is typically done in a fixed environment or network, a notion of \emph{transference} is needed for practical implementation in which the learned policy must maintain strong performance even as the network reconfigures over time.

In this section, we verify the permutation equivariance property of the Agg-GNN architecture with both synchronous and asynchronous aggregation sequences. Moreover, we demonstrate the same permutation equivariance property for the general decentralized resource allocation problem in \eqref{eqn:prob} and establish an optimality of an Agg-GNN policy across permutations of the wireless network, thus facilitating transference of the proposed resource allocation policy across varying network topologies of similar density.

To study the permutation equivariance in wireless networks, recall that the graph $\bbH$ together with input signal $\bbx$ are drawn randomly from a joint distribution $m(\bbH, \bbx)$ and the output is written as \eqref{eqn:output}. Consider that as the underlying network changes in topology, the joint stochastic process distribution is then given by a transformed distribution $\hat{m}(\hat\bbH, \hat\bbx)$ of the stochastic process $\{ \hat\bbH(t), \hat\bbx(t)\}_{t\in\mbZ}$, which is also assumed stationary. With a similar way to \eqref{eqn:local_info}, the history information $\hat\ccalH$ can be formulated as
\begin{align} \label{eqn:permu_local_info}
\hat\ccalH_{i}(t):=\bigcup_{k=0}^{K-1} \Big\{ [\hat\bbH(t-k+1)]_{jj'},& [\hat\bbx(t-k)]_{j'} \mid  \\
&  j\in\hat\ccalN_{i}^{k-1}(t), j'\in\hat\ccalN_{i}^{k }(t) \Big\}, \nonumber
% \ccalH_{i}(t):=\bigcup_{k=0}^{\lceil\frac{K}{n_{ex}} \rceil-1} \bigcup_{n=1}^{n_{ex}} \Big\{ [\bbH(t-k)]_{jj'},& [\bbx(t-k)]_{j'} \mid  \\
% &  j\in\ccalN_{i}^{n-1}(t), j'\in\ccalN_{i}^{n}(t) \Big\}. \nonumber
\end{align}
where the neighboring set can be defined similarly as $\hat\ccalN_{i}(t)=\{ [\hat\bbH(t)]_{ij} \geq \eta_0, j\in \ccalA(t) \}$ and $\hat\ccalN_i^k(t) := \{j'\in\hat\ccalN_j(t-k+1), j\in\hat\ccalN_i^{k-1}(t)\cap \ccalA(t) \}$. 

For wireless network transformations we focus on the case of network permutations. In particular, we define a set of permutation matrices of dimension $m$ as the set of binary matrices defined as
\begin{align}
    \psi =\{\bm{\Pi} \in\{0, 1\}^{m\times m}: \bm{\Pi}\bm{1}=\bm{1},\bm{\Pi}^T\bm{1}=\bm{1} \}.
\end{align}
Applying a permutation matrix $\bm\Pi \in \psi$ to a signal $\bbx$ as $\bm\Pi^T \bbx$ indicates a reordering of elements in the vector. Likewise an application to matrix $\bbH$ as $\bm\Pi^T \bbH \bm\Pi$ indicates a corresponding reordering of columns and rows. In the setting of wireless networks, this can be interpreted as the permutation of locations or labels of the transmitters and paired receivers.

As in Section \ref{sec:prob} we import a function set $\bm\Phi(\ccalH, \bbA)$ with lower dimensional parameters. We first prove that the outputs of the same filter tensor is permutation equivariant, which can be stated as follows.
\begin{proposition}
\label{prop:permute-gnn}
Consider graphs $\bbH$ and $\hat\bbH$ together with signals $\bbx$ and $\hat\bbx$, we have $\hat\bbH=\bm\Pi^T \bbH \bm\Pi$ and $\hat\bbx=\bm\Pi^T \bbx$ for some permutation matrix $\bm\Pi$. The sparsifying matrix $\bbQ$ is also permuted accordingly, which can be written as $\hat\bbQ=\bm\Pi^T \bbQ \bm\Pi$. The output of the Agg-GNN with filter $\bbA$ to the pairs $(\bbH,\bbx)$ and $(\hat\bbH,\hat\bbx)$ are such that:
\begin{gather}
\bm\Phi(\hat\ccalH,\bbA)=\bm\Pi^T \bm\Phi(\ccalH,\bbA).
\end{gather}
\end{proposition}

%In the following proposition, we further establish the permutation equivariance property with asynchronous aggregation sequence generation.
%
%\begin{proposition}
%\label{prop:3}
%Consider graphs $\bbH$ and $\hat\bbH$ together with signals $\bbx$ and $\hat\bbx$ in the  asynchronous setting, we have $\hat\bbH=\bm\Pi^T \bbH \bm\Pi$ and $\hat\bbx=\bm\Pi^T \bbx$ for some permutation matrix $\bm\Pi$. Furthermore, the sparsifying matrix $\bbQ$ and activation matrix $\bbM$ are also permuted accordingly, which can be written as $\hat\bbQ=\bm\Pi^T \bbQ \bm\Pi$ and $\hat\bbM=\bm\Pi^T \bbM \bm\Pi$. The output of the Agg-GNN with filter $\bbA$ to the pairs $(\bbH,\bbx)$ and $(\hat\bbH,\hat\bbx)$ are such that:
%\begin{gather}
%\bm\Phi(\hat\ccalH^0,\bbA)=\bm\Pi^T \bm\Phi(\ccalH^0,\bbA).
%\end{gather}
%\end{proposition}
%\begin{proof}
%The permuted equivalent transmission matrix in the asynchronous setting can be denoted as 
%\begin{gather}
%\hat\bbH(t) \hat\bbM(t)=\bm\Pi^T \bbH(t) \bm\Pi \bm\Pi^T \bbM(t) \bm\Pi =\bm\Pi^T \bbH(t)\bbM(t) \bm\Pi.
%\end{gather}
%These indicate that this linear sparsification operation of the channel matrix in the asynchronous setting can be represented by a general channel transmission matrix and the proof process in Proposition \ref{prop:1} still holds for this conclusion.
%\end{proof}

This proposition states the inherent permutation equarvariance property of the Agg-GNN due to the equivariant manner in which the aggregation sequence is formed and the resulting convolution structure. %We further notice that this permutation equavariance does not need the assumption to hold. 
These results imply that an appropriately permuted output is obtained from a permuted input.

This permutation equivariance is not only a valuable structure for the parametrization to hold, but is moreover a fundamental property of the wireless resource allocation problem itself. We may then study the effect that a permutation brings to the optimal resource allocation problem. In order to do so, we first state the following assumptions that $u_0$ and $\bbu(\bbr)$ are permutation invariant while $\bbf$ is permutation equivariant, which can be written explicitly as follows.
\begin{assumption}
\label{assum}
The utility function $u_0$ and constraint function $\bbu(\bbr)$ are permutation invariant, while reward function $\bbf$ is permutation equivariant, i.e. $\forall \bm\Pi \in \psi$,
\begin{align}
    &u_0(\bm\Pi^T \bbr)=u_0(\bbr)\\
    &\bbu(\bbr)\geq \bm 0 \leftrightarrow \bbu(\bm\Pi^T \bbr)\geq \bm 0\\
    & \bbf(\hat\bbP,\hat\bbH, \hat\bbx)=\bm\Pi^T \bbf(\bbP,\bbH,\bbx),
\end{align}
where $\hat\bbH=\bm\Pi^T\bbH\bm\Pi$ and $\hat\bbx=\bm\Pi^T \bbx$ are channel state and node state permutations respectively, while $\hat\bbP=\bm\Pi^T \bbP$ is a resource allocation permutation.
\end{assumption}
The above assumption means that if the nodes in the network are reordered, the utility function stays the same and the constraint functions are all satisfied as well. This can be realized by carefully designing $u_0$ and $\bbu$, but still holds for many common cases of utilities and constraints. Moreover, the reward function is reordered accordingly to the permuted nodes---a property held for common reward functions, e.g., link capacity. We note that the examples given in Section \ref{sec:eg} all satisfy this assumption. 

Under this assumption, we may establish that the optimal, unparameterized, resource allocation strategy of problem \eqref{eqn:prob} is also permutation equivariant. We state this result in the following proposition from \cite{eisen2020optimal}.

\begin{proposition}
\label{prop:permute-policy}
Consider the wireless resource allocation problem in \eqref{eqn:prob} for a wireless network given by the state distribution $m(\bbH,\bbx)$, where respective functions satisfy Assumption \ref{assum}. Further consider a wireless network permuted by some matrix $\bbPi \in \psi$, given by the probability distribution $\hat{m}(\hat\bbH,\hat\bbx)$ of the stochastic process $\{ \hat\bbH(t), \hat\bbx(t)\}_{t\in\mbZ}$, where
 $\hat\bbH=\bm\Pi^T\bbH\bm\Pi$, $\hat\bbx=\bm\Pi^T\bbx$, and
\begin{align}
\label{eqn:Hpermute}
    \hat{m}(\hat\bbH,\hat\bbx)=\hat{m}(\bm\Pi^T\bbH\bm\Pi, \bm\Pi^T\bbx)=m(\bbH,\bbx).
\end{align}
Further assume there exists a policy that is equivariant to $\bbPi$, i.e.
\begin{align}
    \hat\bbP(\hat\ccalH)=\bm\Pi^T \bbP(\ccalH).
\end{align}
Then, the optimal resource allocation policy $\bbP^*(\ccalH)$ is permutation equivariant, and thus satisfies:
\begin{align}
  \label{eqn:u}& u_0(\hat\bbr)=u_0(\bbr), \quad \bbu(\hat\bbr)\geq \bm{0} \leftrightarrow \bbu(\bbr)\geq \bm{0}.\\
   & \label{eqn:P}\bbP^*(\hat\ccalH)=\bm\Pi^T\bbP^*(\ccalH).
\end{align}
\end{proposition}

From the above proposition, we can see that the resource allocation strategy follows the permutation of nodes in the network, which is consistent with the intuition of how resource allocation should adapt to structural changes in the network. With the propositions brought out above, we can state the theorem from \cite{eisen2020optimal} as follows.
\begin{theorem}
\label{thm:trans}
Consider the wireless resource allocation problem in \eqref{eqn:opt} for a wireless network given by the state distribution $m(\bbH,\bbx)$, where respective functions satisfy Assumption \ref{assum} and the parametrization is given by an Agg-GNN as defined in \eqref{eqn:output}. Further consider a wireless network permuted by some matrix $\bbPi \in \psi$, given by a state distribution $\hat{m}(\hat\bbH,\hat\bbx)$, where
 $\hat\bbH=\bm\Pi^T\bbH\bm\Pi$, $\hat\bbx=\bm\Pi^T\bbx$, and
\begin{align}
\label{eqn:Hpermute1}
    \hat{m}(\hat\bbH,\hat\bbx)=\hat{m}(\bm\Pi^T\bbH\bm\Pi, \bm\Pi^T\bbx)=m(\bbH,\bbx).
\end{align}
Then, the solutions for \eqref{eqn:opt} $\bbA^*$ and $\hat\bbA^*$ under $\hat{m}(\hat\bbH,\hat\bbx)$ and $m(\bbH,\bbx)$ respectively satisfy
\begin{align}
    \hat\bbA^*=\bbA^*.
\end{align}
\end{theorem}

Theorem \ref{thm:trans} states that if two networks are permutations of each other, they have the same optimal Agg-GNNs. Thus, an Agg-GNN trained on a network with state distribution $m(\bbH,\bbx)$ can be transferred to one with state distribution $\hat{m}(\hat\bbH,\hat\bbx)$ without loss of optimality. While the graph $\hat\bbH$ and signal $\hat\bbx$ that compose the history information set $\hat\ccalH$ are different as they follow another distribution, the implemented filter tensor $\bbA$ keeps the same. We call this \emph{transferability} of the trained Agg-GNN. It is important to state this theorem as large scale networks can be seen as permutations of each other. This notion of transference is further explored numerically in Section \ref{sec:sim}.

%%%%%%%%%%%%%%%%%%%%%%%%%%%%%%%%%%%%%%%%%%%%%%%%%%%%%%%%%%%%%%%%%%%%%
%%%%%%%%%%%%%%%%%%%%%%%%%%%%%%%%%%%%%%%%%%%%%%%%%%%%%%%%%%%%%%%%%%%%
%%%   S  E  C  T  I  O  N   %%%%%%%%%%%%%%%%%%%%%%%%%%%%%%%%%%%%%%
%%%%%%%%%%%%%%%%%%%%%%%%%%%%%%%%%%%%%%%%%%%%%%%%%%%%%%%%%%%%%%%%%%%%
%%%%%%%%%%%%%%%%%%%%%%%%%%%%%%%%%%%%%%%%%%%%%%%%%%%%%%%%%%%%%%%%%%%%%
\section{Primal-Dual Training}
\label{sec:primal}
The optimal Agg-GNN for allocating resources in the wireless network is specified by the optimal filter tensor $\bbA^*$ given in \eqref{eqn:opt}. To solve the constrained optimization problem, we convert the constrained form in \eqref{eqn:opt} to the so-called Lagrangian function, i.e., 
\begin{align}\label{eqn:lagrange}
    \nonumber\mathcal{L}(\bbA, \bbr, \bm{\lambda}, &\bm{\mu})=u_0(\mathbf{r})+ \\
    &\bm\lambda^T  \left[\mathbb{E}\left[\bbf\left(\bm{\Phi}(\ccalH,\bbA), \bbH,\bbx\right) \right]-\bbr \right]+\bm{\mu}^T \bbu(\bbr),
\end{align}
where $\bm\lambda, \bm\mu \geq \bb0$ are introduced as the dual variables that penalize constraint violation in \eqref{eqn:lagrange}. The resulting dual optimization problem of \eqref{eqn:opt} consists of maximizing and minimizing $\ccalL$ with respect to the primal and dual variables, respectively, i.e.,
\begin{equation}\label{eqn:dual}
[\hbA^*, \hbr^*, \bblambda^*, \bbmu^*] = \min_{\bblambda, \bbmu \geq \bb0} \max_{\bbA, \bbr} \ccalL(\bbA, \bbr, \bm\lambda, \bm\mu).
\end{equation}
Observe in \eqref{eqn:dual} that the optimal filter tensor of the dual problem $\hbA^*$ and associated expected rewards $\hbr^*$ are found as the saddle point of the Lagrangian function with dual variables $\bm\lambda$ and $\bm\mu$. For sufficiently dense parametrizations it can be shown that dual-optimal filter tensor is close to that of the original constrained problem in \eqref{eqn:opt} \cite{eisen2019learning}.

The primal-dual method often employed is to alternatively update primal and dual variables with gradient ascent and descent respectively. Let $\tau$ denote an iteration index and $\epsilon > 0$ as the step-size. Due to the asynchronous activation patterns of nodes, inactive nodes cannot utilize the current parameter tensor $\bbA(\tau)$. The centralized learner keeps and updates $\bbA(\tau)$, while we further define the local copy at node $i$ as $\bbA_i(\tau)$. We store all the local copies together as $\mathbb{A}(\tau):=\{\bbA_i(\tau)\}_{i=1}^m$. The local copies are expressed are:
\begin{equation}\label{eqn:lcl_A'}
\bbA_i(\tau) := \begin{cases}
\bbA(\tau) & \text{if } i \in \ccalA(\tau), \\
\bbA_i(\tau-1) & \text{if } i \notin \ccalA(\tau).
\end{cases}
\end{equation}

% \begin{align}
%   \label{eqn:prim1} \mathbf{r}_{k+1}&=\mathbf{r}_k+\epsilon [\nabla_{\mathbf{r}} u_0(\mathbf{r}_k)+ \nabla_{\mathbf{r}} \bbu(\mathbf{r}_k)\bm{\mu}_k - \bm{\lambda}_k],\\
%   \label{eqn:prim2} \bbA(u+1) &= \bbA(u) + \epsilon \left[\nabla_{\bbA}\E\left[\sum_{i=1}^m \mathbf{f}_i(\mathbf{\Phi}(\ccalH_{it};\bbA);\ccalH_{it}) \right]\right]\bm{\lambda}_k
% \end{align}
The primal updates are obtained by gradient ascent updates on the Lagrangian function,
\begin{align}
   \label{eqn:prim1} \bbr(\tau+1)&=\bbr(\tau)+\epsilon [\nabla_{\bbr} u_0(\bbr(\tau))+ \nabla_{\bbr} \bbu(\bbr(\tau))\bm{\mu}(\tau) -  \bm\lambda(\tau)],\\
   \label{eqn:prim2} \bbA(\tau+1) &= \bbA(\tau) + \epsilon \left[\nabla_{\bbA}\mathbb{E}\left[\bbf\left(\bm\Phi(\ccalH,\mathbb{A}), \bbH,\bbx\right) \right]\right] \bm\lambda(\tau).
\end{align}

Likewise, the dual variables are updated by performing gradient descent iterations on the Lagrangian function,
\begin{align}
   \label{eqn:dual1} \bm\mu(\tau+1) &= [\bm\mu(\tau) -\epsilon\bbu(\bbr(\tau))]^+,\\
  \label{eqn:dual2}  \bm\lambda(\tau+1)& = \bm\lambda(\tau) - \epsilon\left[\mathbb{E}\left[\bbf\left(\mathbf{\Phi}(\ccalH,\mathbb{A}), \bbH,\bbx\right) \right]-\bbr(\tau)\right].
%   \\
%   \label{eqn:dual3} \bm\lambda(u+1) & =\bm\lambda(u)- \epsilon\left[\bbf\left(\mathbf{\Phi}(\ccalH,\bbA), \bbH,\bbx\right) -\bbr(u)\right].
\end{align}

As we can notice, \eqref{eqn:prim2} and \eqref{eqn:dual2} cannot be computed without explicit knowledge of distribution $m(\bbH, \bbx)$. This can be resolved by using stochastic updates by sampling a realization $(\bbH(\tau), \bbx(\tau))$ and update according to:
\begin{align}
\nonumber \bbA(\tau+1)& = \bbA(\tau) + \\\label{eqn:prim3} &\epsilon \left[\nabla_{\bbA} \bbf\left(\mathbf{\Phi}(\ccalH(\tau),\mathbb{A}(\tau)), \bbH(\tau),\bbx(\tau) \right) \right]  \bm\lambda(\tau),\\
    \label{eqn:dual3} \bm\lambda(\tau+1) & =\bm\lambda(\tau)- \epsilon\left[\bbf\left(\mathbf{\Phi}(\ccalH(\tau),\mathbb{A}(\tau)), \bbH(\tau),\bbx(\tau)\right) -\bbr(\tau)\right].
\end{align}
To implement \eqref{eqn:dual3}, we can use the observed outcome $\bbf\left(\mathbf{\Phi}(\ccalH(\tau),\mathbb{A}(\tau)), \bbH(\tau),\bbx(\tau)\right)$ directly without the need to know the explicit model of function $\bbf$. However, for \eqref{eqn:prim3} we need the gradient of $\bbf$ which cannot be observed from the system. We mimic the randomized policies employed in policy gradient methods\cite{sutton2000policy} and see $\bm\Phi(\ccalH, \mathbb{A})$ as a random variable with probability distribution $\bm\Psi(\ccalH, \mathbb{A})$. The gradient therefore can be rewritten as
\begin{align}
    \nonumber \nabla_\bbA \mathbb{E}_{\bm\Phi}&\left[ \bbf\left( \bm\Phi(\ccalH, \mathbb{A} ),\bbH, \bbx \right) \right] =\\ &\quad \mathbb{E}_{\bm\Phi}\left[ \bbf\left( \bm\Phi(\ccalH, \mathbb{A}),\bbH, \bbx \right) \nabla_\bbA \log\bm\Psi(\ccalH, \mathbb{A})^T \right].
\end{align}
The unknown gradient of $\bbf$ is replaced with the expectation of the gradient of $\bm\Psi$, which can be set by assuming a common distribution $\bm\Psi(\ccalH, \mathbb{A})$. With this gradient estimation implemented, $\bbA$ can be updated as
\begin{align}
    \nonumber \bbA(\tau+1) = \bbA(\tau) + &\epsilon [ \bbf\left(\mathbf{\Phi}(\ccalH(\tau), \mathbb{A}(\tau)), \bbH(\tau),\bbx(\tau) \right)  \\\label{eqn:prim4} &\quad \nabla_{\bbA} \log\bm\Psi(\ccalH(\tau), \mathbb{A}(\tau))^T   \bm\lambda(\tau) ].
\end{align}

% and compute with an approximated gradient as
% \begin{align}
%   \nonumber\bbA(\tau+1) &= \bbA(\tau) + \\\label{eqn:prim4} &\quad\epsilon \left[\tilde\nabla_{\bbA} \bbf\left(\mathbf{\Phi}(\ccalH(\tau),\bbA(\tau)), \bbH(\tau),\bbx(\tau) \right) \right]  \bm\lambda(\tau).
% \end{align}
% \begin{align}
%   \label{eqn:dual1} \bm{\mu}_{k+1} &= [\bm{\mu}_k -\epsilon\bbu(\mathbf{r}_k)]^+,\\
%   \label{eqn:dual2}  \bm{\lambda}_{k+1}& =\bm{\lambda}_k - \left[\E\left[\sum_{i=1}^m \mathbf{f}_i(\mathbf{\Phi}(\ccalH_{it};\bbA);\ccalH_{it}) \right]-\mathbf{r}_k\right]\\
%   \label{eqn:dual3} \bm{\lambda}_{k+1} & =\bm{\lambda}_k- \left[\sum_{i=1}^m \mathbf{f}_i(\mathbf{\Phi}(\ccalH_{ik};\bbA(u));\ccalH_{ik}) -\mathbf{r}_k\right].
% \end{align}

%\red{From \eqref{eqn:dual2} to \eqref{eqn:dual3}, we employ the stochastic update method by approximating the expectation with samplings.} \blue{Both (19) and (21) need to be approximated with stochastic updates. Explain this notion in a bit greater detail (motivated by no model for $\bbf$) and re-present the stochastic versions of these updates (e.g. (22) and the stochastic version of (19))}

The detailed algorithm for {the primal-dual training of the } Agg-GNN is shown in Algorithm \ref{alg:Agg-GNN}. In Step \ref{step:generate}, each node generates its own local sequence following Algorithm \ref{alg:Sequence}. Together with current states, each node can compute the allocation strategy and probe the system $\bbf\left(  \bm{\Phi}(\ccalH(\tau),\mathbb{A}(\tau)), \bbH(\tau),\bbx(\tau)\right)$. In Step \ref{step:p-d}, the primal-dual gradient updates are performed. The process is repeated until convergence.

\begin{remark}
\label{remark_c_training}\normalfont
%\blue{add remark about centralized training vs. decentralized execution/inference}
We note here that the training of the allocation function is done jointly at all nodes, which means that only the data known by all the nodes are used to train the network. All the nodes are working collaboratively to maximize the same global objective and in the mean time sharing the network parameters. The execution of the allocation strategy is decentralized at each node in the meanwhile.  This framework is often employed in many existing works due to its benefits for resource cost and stability. The decision of each node is still different due to different aggregation input sequence on each node. 

\end{remark}

\begin{remark}
\label{remark_training_scale}\normalfont
We stress that the time index used in the primal-dual training process $\tau$ does not need to be the same as the execution time index $t$. That is, the training process can be performed offline using any sequence of state samples from previous experience or previously collected and stored information. We use the separate indices $\tau$ and $t$ to emphasize this difference in time.
\end{remark}

\begin{algorithm}
\caption{Primal-Dual Training Method}
\label{alg:Agg-GNN}
\begin{algorithmic}[1]
\FOR{$\tau\in\mbZ$ }
\STATE{Transmitters generate aggregation sequence  and decide their resource levels $\bm{\Phi}(\ccalH(\tau), \bbA(\tau))$ w. Alg. \ref{alg:Sequence} based on current channel states and parameter set. }\label{step:generate}
%\STATE{Node $i$ compute local action with policy
%$\bm{\phi}(\ccalH_{i}(\tau);\bbA_i(t))$}
\STATE{Observe rate feedback 
$\bbf\left(  \bm{\Phi}(\ccalH(\tau), \bbA(\tau)), \bbH(\tau),\bbx(\tau)\right)$ }
\STATE{Update primal and dual variables as \eqref{eqn:prim1}-\eqref{eqn:dual2}
\begin{align*}
&\bbr(\tau)+\epsilon [\nabla_{\bbr} u_0(\bbr(\tau))+ \nabla_{\bbr} \bbu(\bbr(\tau))\bm{\mu}(\tau) -  \bm\lambda(\tau)],\\
&[\bm\mu(\tau) -\epsilon\bbu(\bbr(\tau))]^+,\\
&\bm\lambda(\tau)- \epsilon \left[\bbf\left(\mathbf{\Phi}(\ccalH(\tau),\mathbb{A}(\tau)), \bbH(\tau),\bbx(\tau)\right) -\bbr(\tau)\right],\\
&\bbA(\tau) +  
\epsilon [ \bbf\left(\mathbf{\Phi}(\ccalH(\tau), \mathbb{A}(\tau)), \bbH(\tau),\bbx(\tau) \right)\\ &\qquad\qquad\qquad \nabla_{\bbA} \log\bm\Psi(\ccalH(\tau), \mathbb{A}(\tau))^T   \bm\lambda(\tau) ]
\end{align*}
}\label{step:p-d}
\ENDFOR
\end{algorithmic}
\end{algorithm}

\section{Numerical Experiments}
\label{sec:sim}
In this section, we provide a numerical study of the performance of the proposed Agg-GNN parametrization for decentralized power control problem among $m$ transmitters over an AWGN channel with interference as presented in Section \ref{sec:dynpowAWGN}. With utility function set as the sum-rate capacity, constraint function can be set as the maximum total power budget $P_{max}$. The complete problem can be formulated as
\begin{align}
\label{eqn:prob_sim}
\bbr^*&=\max_{\bbp} \sum_{i=1}^m r_i\\
   s.t.\quad \nonumber &\bbr=\mathbb{E}\left[  \log\left(1+\frac{|h_{ii}(t)|^2 p_i(\ccalH_{i}(t))}{1+ \sum\limits_{j\in\ccalN_{it}} |h_{ij}(t)|^2 p_j(\ccalH_{j}(t))}\right) \right],\\
   \nonumber & \mathbb{E}[\bm{1}^T\bbp]\leq P_{max},\quad p_i(\ccalH_{i}(t))\in \{0,p_0\}.
\end{align}
The Agg-GNN is trained in a model-free manner using the primal-dual policy gradient method presented in Algorithm \ref{alg:Agg-GNN} in a variety of representative wireless network scenarios. In all cases, we verify its performance by comparing against a set of both model-based and model-free baseline power allocation methods. 

\subsection{Synchronous setting}\label{sec_nn1}
We begin by studying the wireless ad-hoc networks where each transmitter has a unique receiver, i.e. $m=n$, and assume that nodes operate on synchronous clocks, i.e. $\ccalA(t) \equiv \{1,\hdots,m\}$. To construct this network, we first drop $m$ transmitters randomly uniformly within the range of $\bba_i\in[-m,m]^2, i=1,2\dots, m$. Each paired receiver is located randomly within $\bbb_i\in [\bba_i-m/4, \bba_i+m/4]^2$. The fading channel state is composed of a large-scale pathloss gain and a random fast fading gain, which can be written as $h_{ij}=h^l_{ij}h^f_{ij}$, $h^l_{ij}=\| \bba_i - \bbb_j \|^{-2.2}$, the real and imaginary part at initial time is $h^{
fr}_{ij}(0), h^{fi}_{ij}(0)\sim \ccalN(0,1)$ as \eqref{eqn:Gau-Mar} indicates. The relativity of channel coefficients is measured by $\delta$ which is set as $0.3$ in this scenario. %$h^f_{ij}\sim \text{Rayleigh}(2)$. 

By employing the algorithm we present in Algorithm \ref{alg:Agg-GNN}, we train an Agg-GNN with $L=10$ hidden layers, each with $F_l=1$ filter with length $K_l=10$ and a standard ReLu nonlinear activation function, i.e. $\bm{\sigma}(\bby)=[\bby]_+$. The final layer normalizes the outputs through a sigmoid function. {The number of total parameters trained therefore is 100 which is invariant to the size of wireless networks.} We compare our algorithm with three existing heuristic methods for solving the problem stated in \eqref{eqn:prob_sim}: 
\begin{itemize}
    \item WMMSE \cite{shi2011iteratively} in a distributed setting with fixed number of iterations per allocation decision. This is a \emph{model-based} approach that assumes knowledge of the capacity function in \eqref{eqn:prob_sim},
    \item Equal power allocation, i.e. assign $P_{max}/m$ to all transmitters,
    \item Random full power allocation, i.e. each transmitter transmits with full power $p_0$ with probability $P_{\max}/(p_0 m)$.
\end{itemize}
In \cite{shi2011iteratively}, the information exchange complexity is controlled by setting a maximum number of iterations in the algorithm. Similarly, in our algorithm Agg-GNN, the complexity of information exchanges is measured by the maximal neighborhood range. To compare these two distributed algorithms, we set the number of iterations and maximal hop size as the same. %\green{WMMSE also implements the delayed information.} 
Finally, we also compare with the existing Selection Graph Neural Network method \cite{eisen2020optimal}, which we stress is a \emph{centralized} implementation. The network is set with $L=10$ hidden layers, each with $F_l=1$ graph filter of length $K_l=10$. The centralized GNN is trained in the same model-free manner as used for the Agg-GNN. 

\begin{figure}[t]
\centering
\centerline{\includegraphics[width=0.5\textwidth]{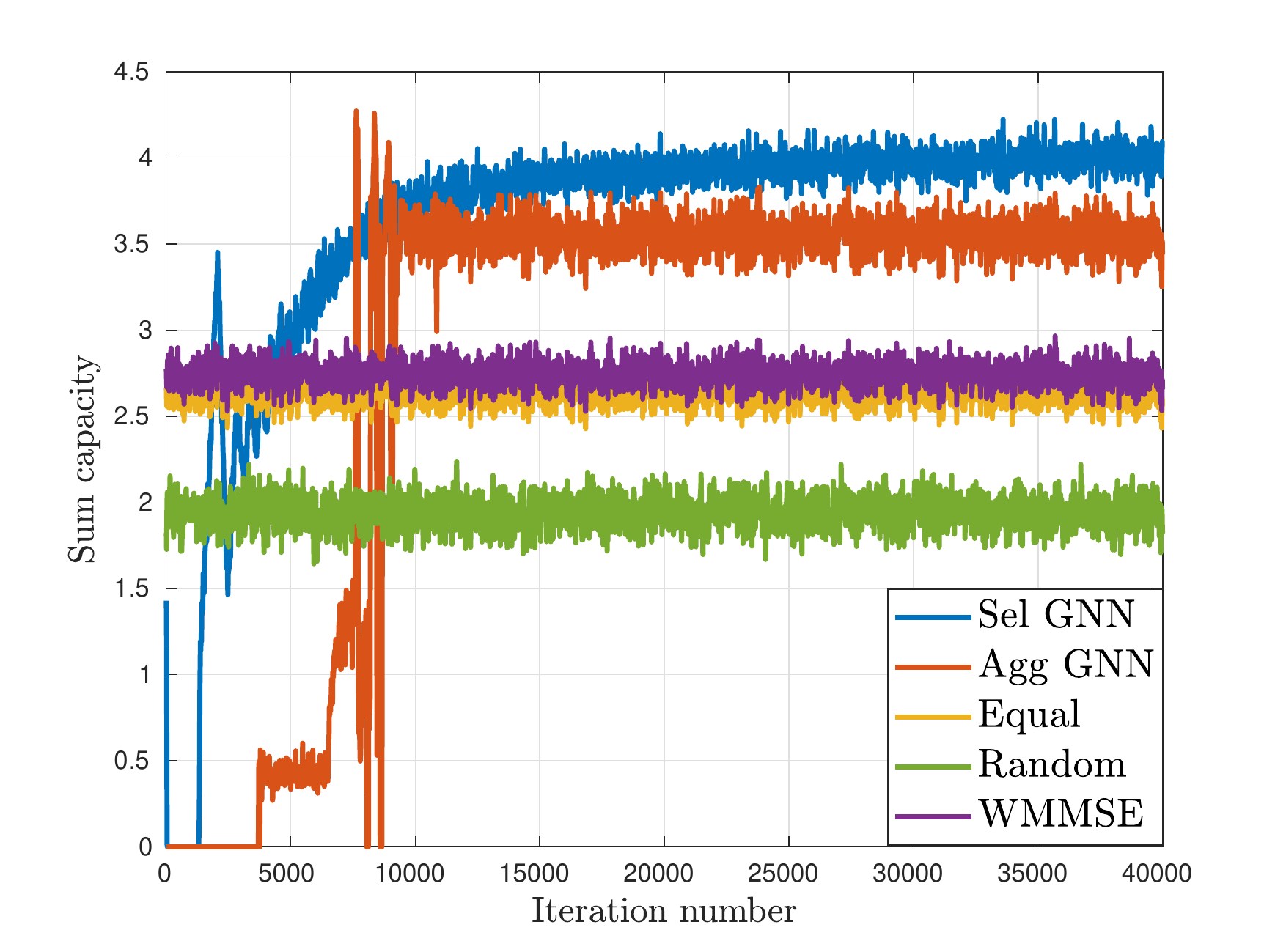}}
\caption{Performance comparison during training for 25 nodes with 5 hops.}
\label{fig:comp25}
\end{figure}

\begin{figure}[t]
\centering
\centerline{\includegraphics[width=0.5\textwidth]{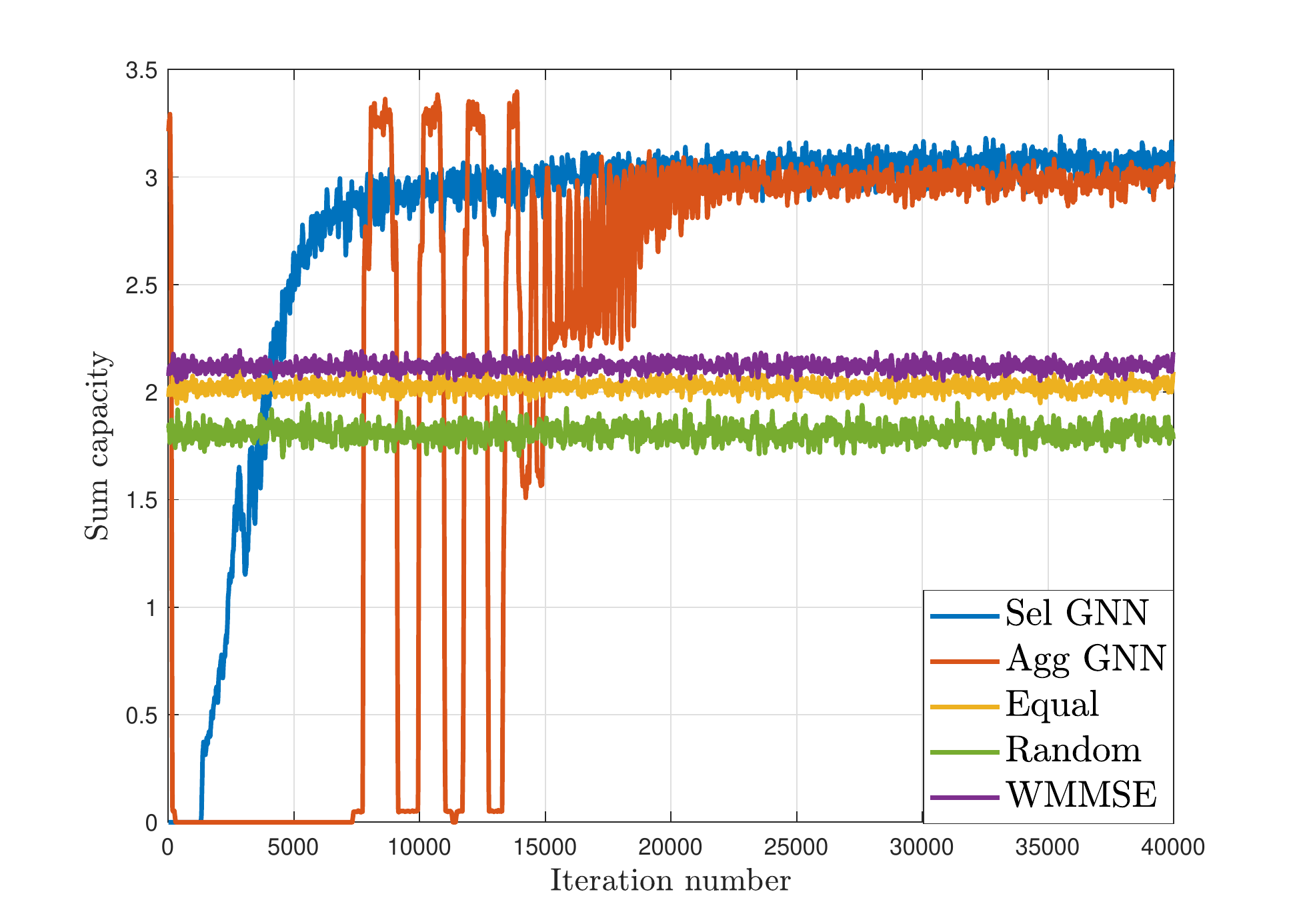}}
\caption{Performance comparison during training for 50 nodes with 6 hops.}
\label{fig:comp50}
\end{figure}

\begin{figure}[t]
\centering
\centerline{\includegraphics[width=0.5\textwidth]{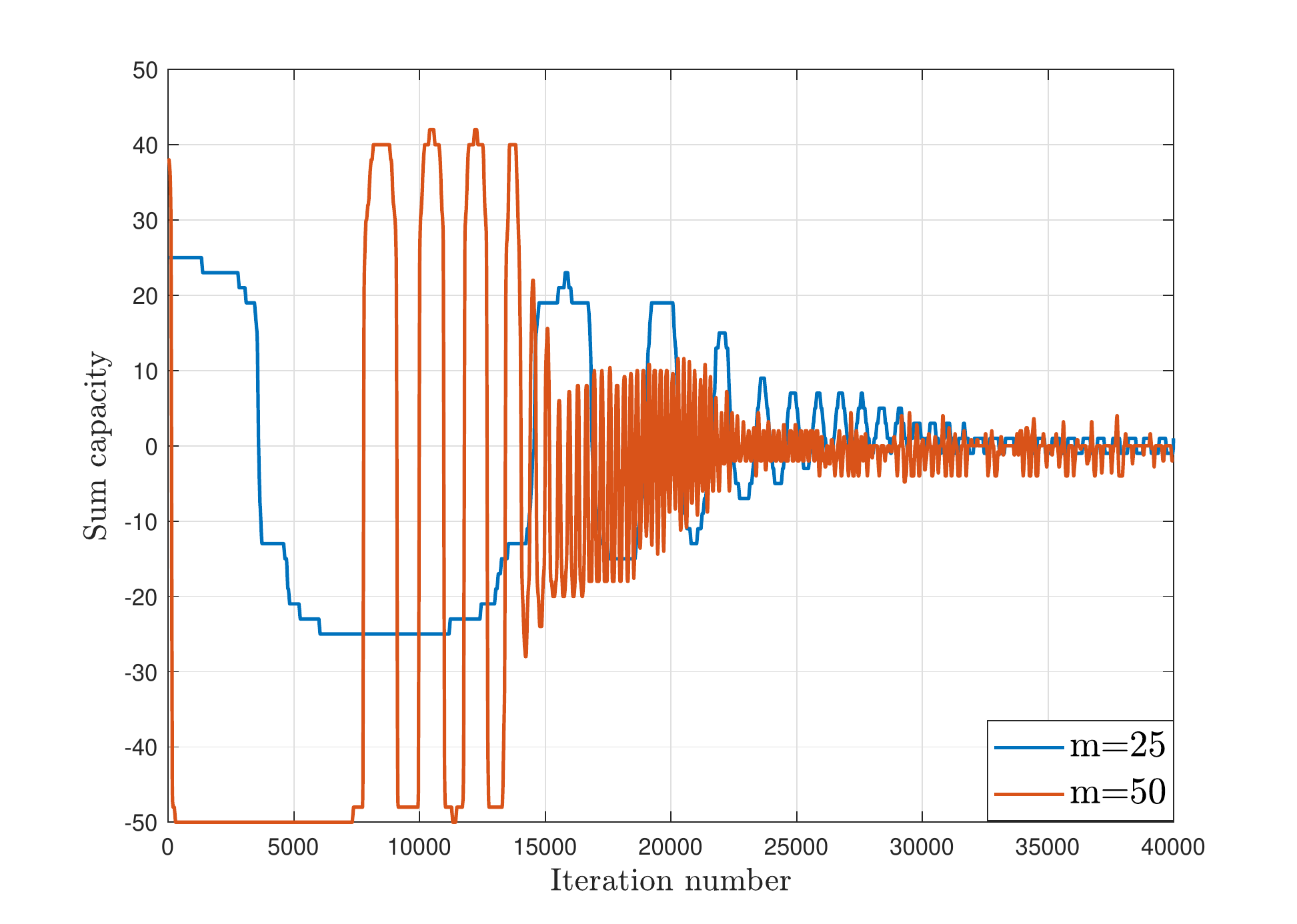} }
\caption{Constraint violation of Agg-GNN during the learning process for 25 nodes and 50 nodes ad-hoc networks. }
\label{fig:ConVio}
\end{figure}

In Figure \ref{fig:comp25} we show the performance through the learning process of the above mentioned algorithms under a medium scale network system with $m=25$ transmitter-receiver pairs. It can be seen that Agg-GNN outperforms all decentralized methods while almost matching the performance of the centralized Sel-GNN method. The primal-dual training process of Agg-GNN converges slower than that of Sel-GNN due to limited local information. While the Agg-GNN performance gain over WMMSE is small, the performance gain is nonetheless achieved in a model-free manner. In Figure \ref{fig:comp50} we show the same comparison with a larger network setting with $m=50$ pairs. Here, the Agg-GNN matches the performance of the Sel-GNN and significantly outperforms all decentralized baseline methods. These results suggest a greater opportunity for gain over existing heuristics in larger network scenarios. To be sure of constraints satisfaction, we further check the constraint violation during the primal-dual training process for these two scenarios of Agg-GNN. The result is shown in Figure \ref{fig:ConVio}, demonstrating proper satisfaction in both network scenarios.

\begin{figure}[t]
\centering
\centerline{\includegraphics[width=0.5\textwidth]{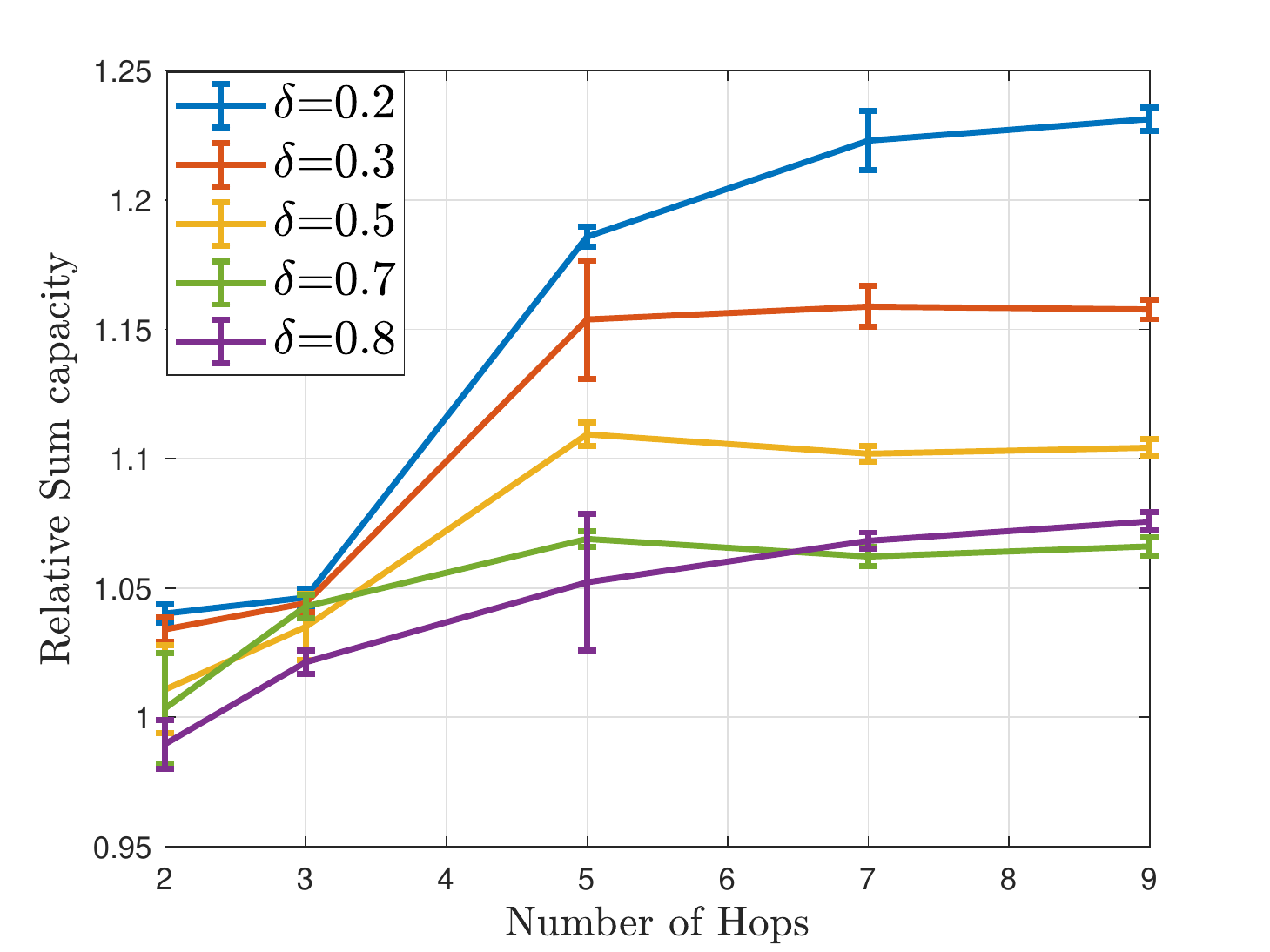}}
\caption{Performance comparison of different hop sizes for 25 nodes ad-hoc networks.}
\label{fig:hopcomp25}
\end{figure}

Next, we study the performance of Agg-GNN with different values of maximal neighborhood range under different coherence time settings. Figure \ref{fig:hopcomp25} shows the final performance of Agg-GNN with respect to the length of aggregation, i.e. $K$. We can see that the performance would converge after a certain aggregation length, which indicates only finite number of information exchanges are needed for a large network. This value tends to indicate the diameter of the network. The coherence time can be reflected by $\delta$ in the channel model \eqref{eqn:Gau-Mar}. As $\delta$ gets smaller, the correlation between each time step becomes stronger. Aggregation information of the same length therefore helps more to make current decision. As is shown in Figure \ref{fig:hopcomp25}, the performance increases as $\delta$ gets smaller and there is greater correlation in delayed channel state information---see Remark \ref{remark_coherence}. The relative sum of capacity indicates the ratio of the sum of capacity achieved by Agg-GNN to that of WMMSE so as to normalize the difference caused by the change of channel distributions.

% \begin{figure}[t]
% \centering
% \centerline{\includegraphics[width=0.5\textwidth]{figures/aggstep50.eps}}
% \caption{Performance comparison of different hop sizes for 50 nodes ad-hoc networks.}
% \label{fig:hopcomp50}
% \end{figure}

\subsection{Asynchronous setting}

We next evaluate the performance of the Agg-GNN and primal-dual learning method for decentralized resource allocation under the asynchronous setting. We model this asynchrony of working and sleeping patterns of different transmitters by considering a collection of $N_{act}$ active subsets denoted as $\{\ccalA_i\}_{i=1}^{N_{act}}$, where $\ccalA_i \subseteq  \{1,2,\hdots,m\}$ for all $i$. In our simulations such subsets are generated randomly. At each time $t$, we randomly draw a set of active nodes $\ccalA(t) \in\{\ccalA_i\}_{i=1}^{N_{act}}$.

\begin{figure}[t]
\centering
\centerline{\includegraphics[width=0.5\textwidth]{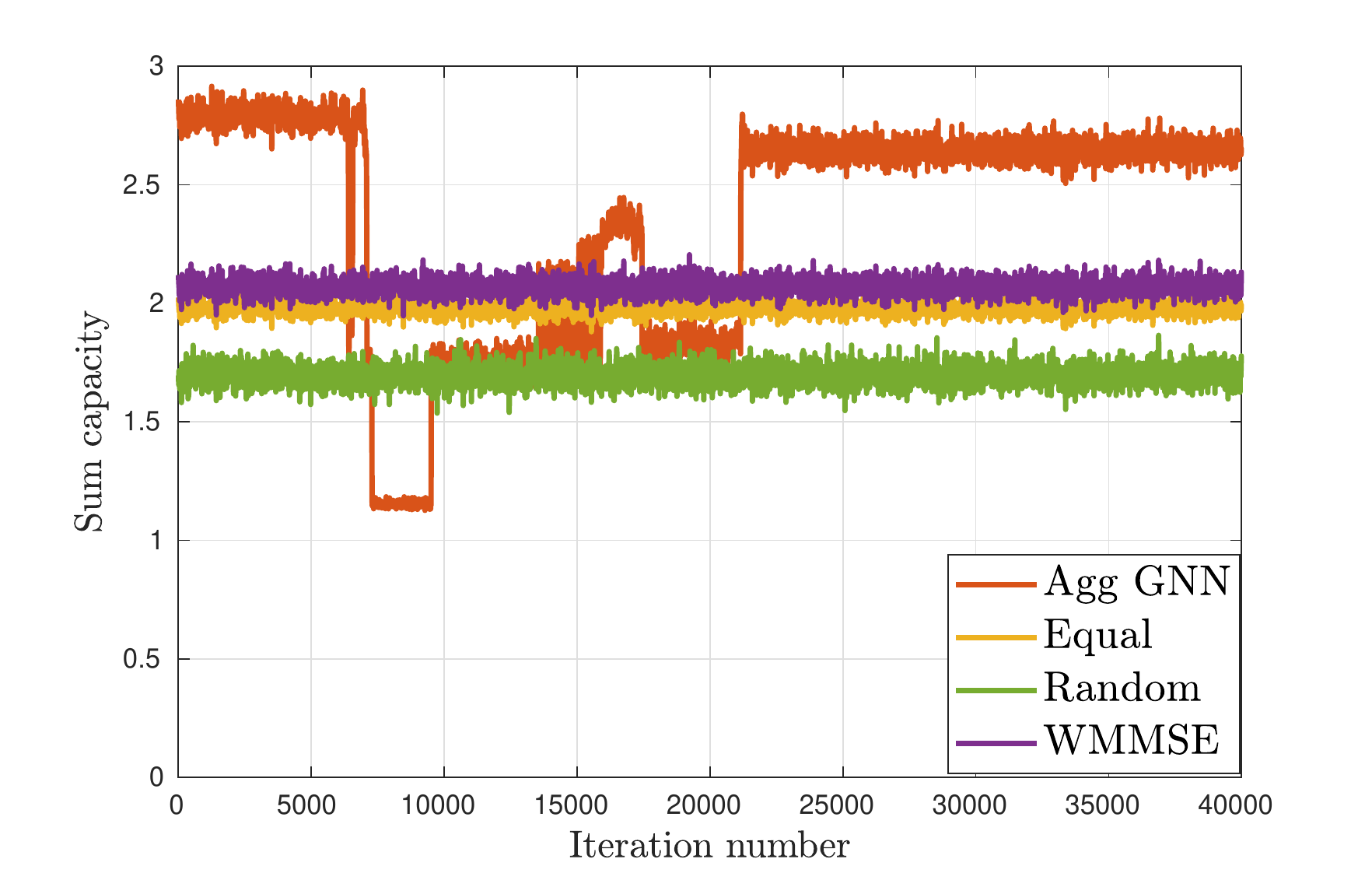}}
\caption{Performance comparison for 50 nodes in an asynchronous setting.}
\label{fig:asyn50}
\end{figure}

Under the asynchronous setting, we demonstrate the performance of the Agg-GNN relative to baseline methods during the primal-dual training process in Figure \ref{fig:asyn50} for 50 nodes setting. We set the  number of active nodes at each time step to be a Poisson distributed random variable with $\lambda=25$. To be consistent with the comparison, here the Equal, Random and WMMSE algorithms are set to update resource allocation actions every $m/\lambda=2$ time slots. Observe that, while the performance degrades relative to the synchronous setting, the Agg-GNN still either meets or exceeds the performance of the non-GNN baseline methods. It can further be observed that, due to the additional noise in random wake patterns, there is greater oscillation in the convergence curves in the asynchronous setting.

%\begin{figure}[t]
%\centering
%\centerline{\includegraphics[width=0.5\textwidth]{figures/asyn25new.eps}}
%\caption{Performance comparison for 25 nodes in an asynchronous setting.}
%\label{fig:asyn25}
%\end{figure}

\subsection{Transference}
\begin{figure}[h]
\centering
\centerline{\includegraphics[width=0.5\textwidth]{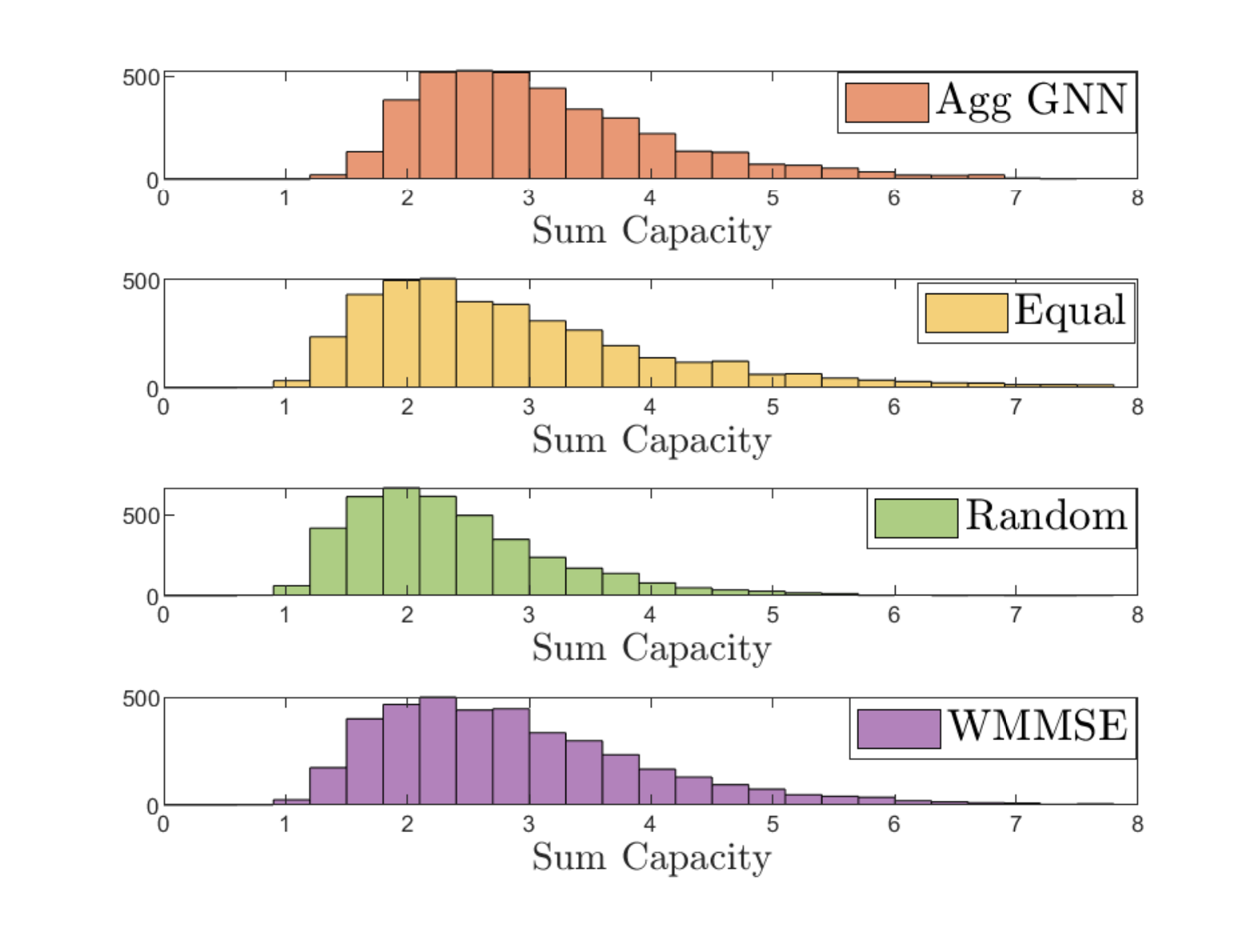}}
\caption{Performance comparison in other randomly drawn networks of equal size with $m=25$.}
\label{fig:trans25}
\end{figure}

\begin{figure}[h]
\centering
\centerline{\includegraphics[width=0.5\textwidth]{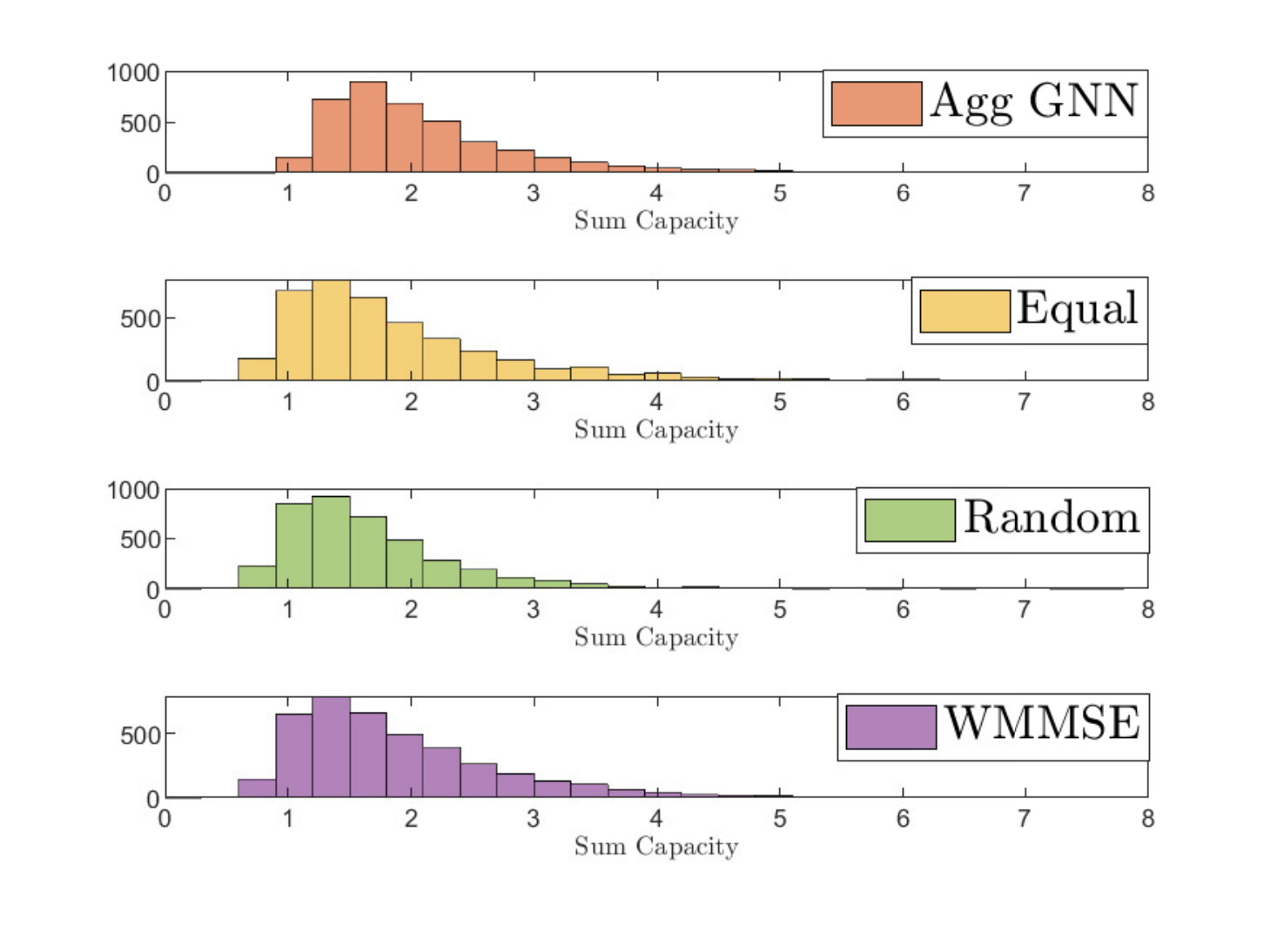}}
\caption{ Performance comparison in other randomly drawn networks of equal size with $m=50$.}
\label{fig:trans50}
\end{figure}

%\blue{I believe we should remove Sel GNN from the figures in this section.}
In this section we study the transference of the learned graph neural network by training the Agg-GNN in a fixed wireless network and observing the performance on new randomly drawn networks of the same or increasing size.  As we have presented in Theorem \ref{thm:trans}, an Agg-GNN remains optimal across permutations of a wireless network. Thus, we may expect that an Agg-GNN trained to exhibit strong performance on a single network should exhibit strong performance on new networks, given that they are close to the original network up to some permutations. Here we investigate the transference capabilities on the randomly drawn networks with equal size and density as shown in Figure \ref{fig:trans25} and Figure \ref{fig:trans50} for networks of size $m=25$ and $m=50$ respectively. The histograms show the empirical distribution of sum-of-rate achieved under randomly generated networks. We can see from the results that the learned Agg-GNN performs well on another network, as suggested by the permutation equivariance of the policy.

\begin{figure}[t]
\centering
\centerline{\includegraphics[width=0.5\textwidth]{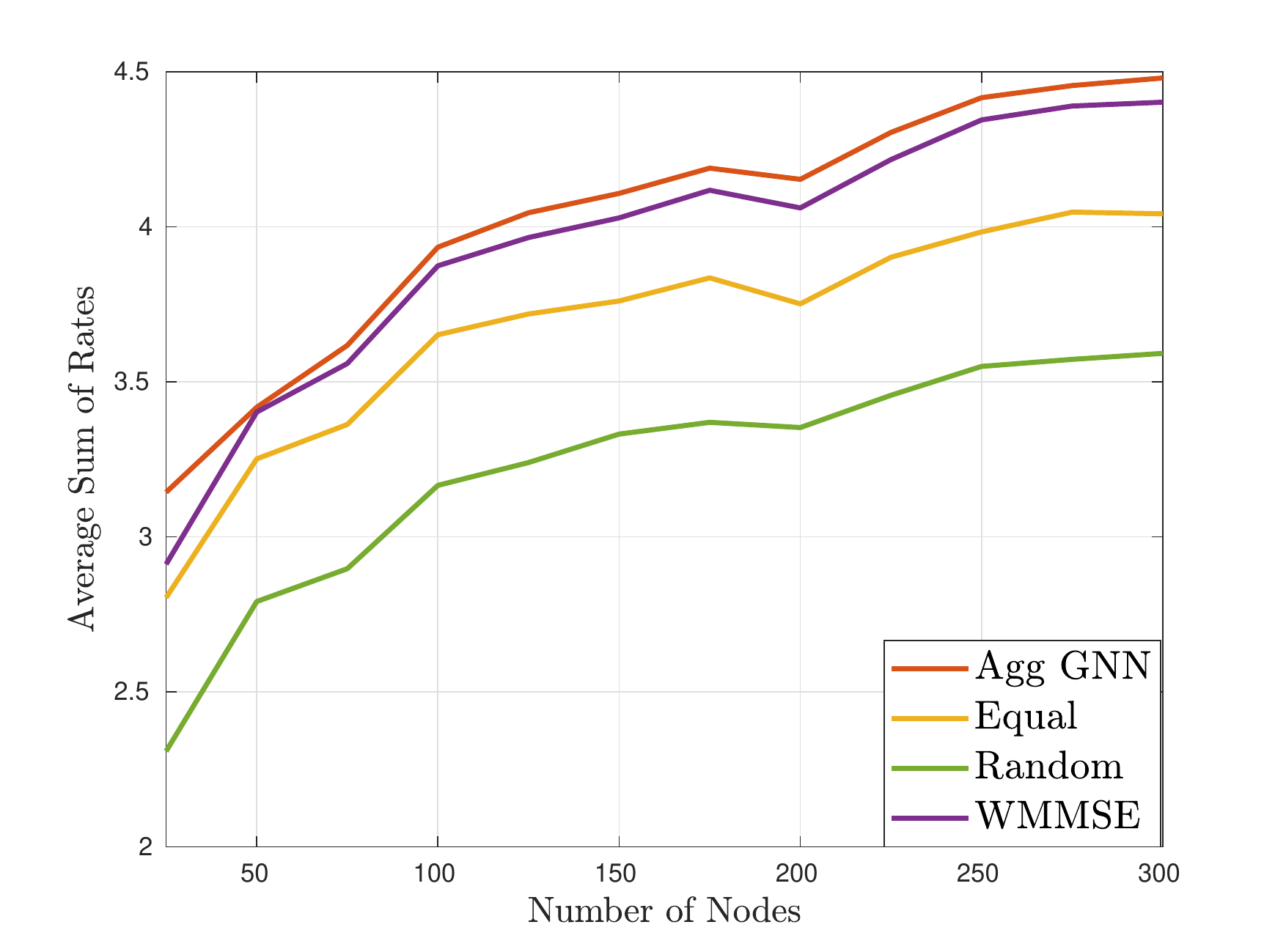}}
\caption{Performance comparison in larger randomly drawn networks of equal density with $m=25$.}
\label{fig:translarger25}
\end{figure}

\begin{figure}[t]
\centering
\centerline{\includegraphics[width=0.5\textwidth]{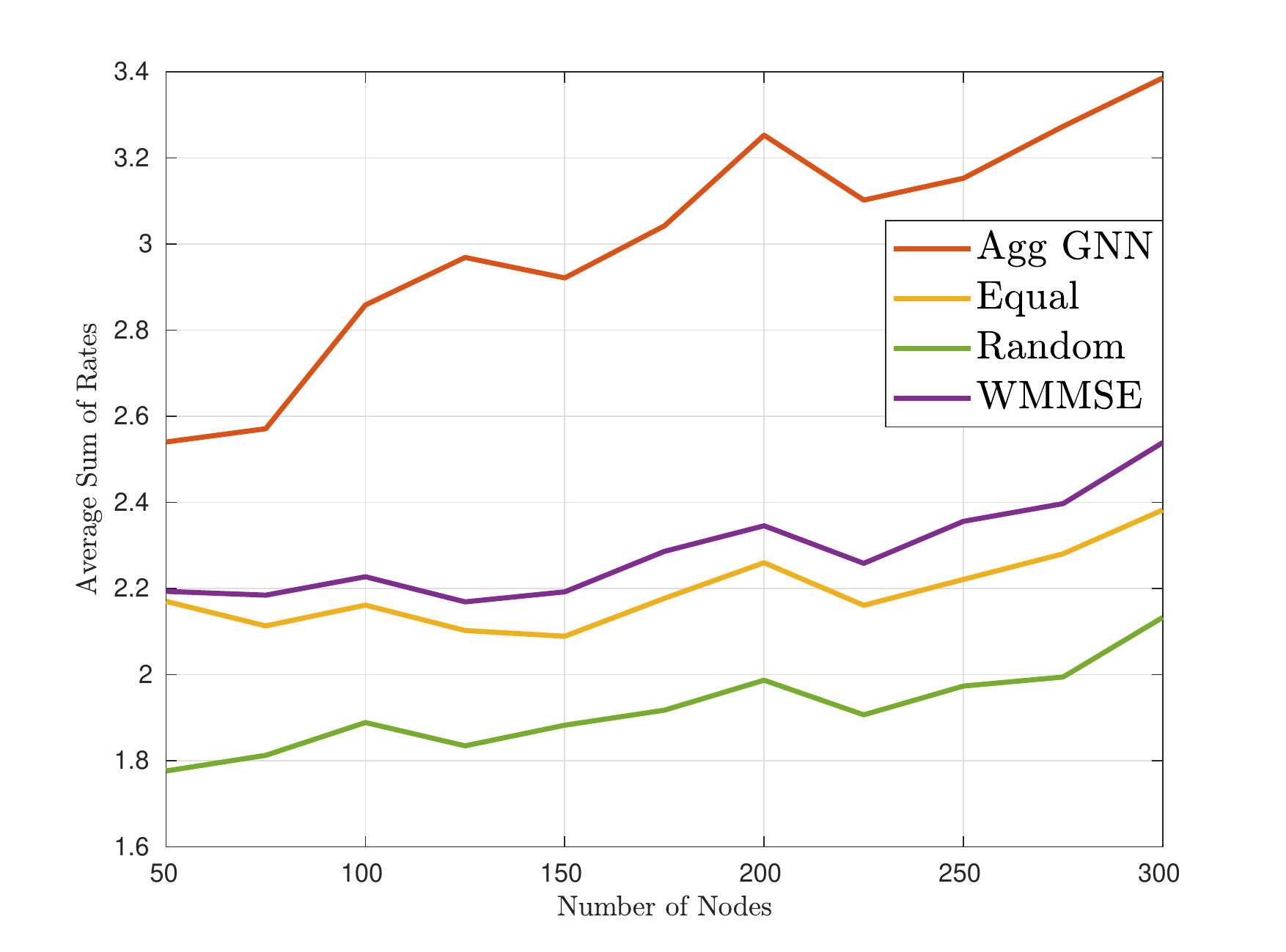}}
\caption{Performance comparison in larger randomly drawn networks of equal density with $m=50$.}
\label{fig:translarger50}
\end{figure}

Another form of transference that is critical for the implementation of learning models in large scale wireless networks is a transference \emph{across scale}, by which the performance of an Agg-GNN is evaluated in new networks with fixed density but varying increasing size. In other words, we train a neural network on a network with $m$ nodes and evaluate on a network with $m'$ nodes. Recall that the Agg-GNN is fully specified by the filter tensor $\bbA$, which acts directly on the aggregation sequence and can therefore be implemented directly on networks (i.e. graphs) of any size. To keep the network density fixed, we drop the transmitters randomly at $\bba_i\in[-\sqrt{mm'},\sqrt{mm'}]^2 $ and its correspondent receiver at $\bbb_i\in[\bba_i-m/4, \bba_i+m/4]^2$. We explore the performance of the learned Agg-GNN in networks of increasing size. Agg-GNNs are trained on networks of size $m=25$ and $m=50$ respectively, and the performances are shown in Figure \ref{fig:translarger25} and Figure \ref{fig:translarger50} in randomly drawn networks of increasing $m'$ relative to heuristic baselines. It can be observed that even with the network size increasing, the original Agg-GNN {with a fixed number of trained parameters} continues to outperform other heuristic methods. {This indicates that Agg-GNNs are more efficient in parameter size than traditional neural networks due to the structural properties imposed in the architecture as we have discussed in Section \ref{sec:prob}.} The demonstrated capability of transferring across scale suggests that, while fully large scale wireless networks may be difficult to access during the training process, it is sufficient to train an Agg-GNN on a representative smaller network given their permutation equivariance and stability properties.

\subsection{Multi-cell interference network}
\begin{figure}[t]
\centering
\centerline{\includegraphics[width=0.5\textwidth]{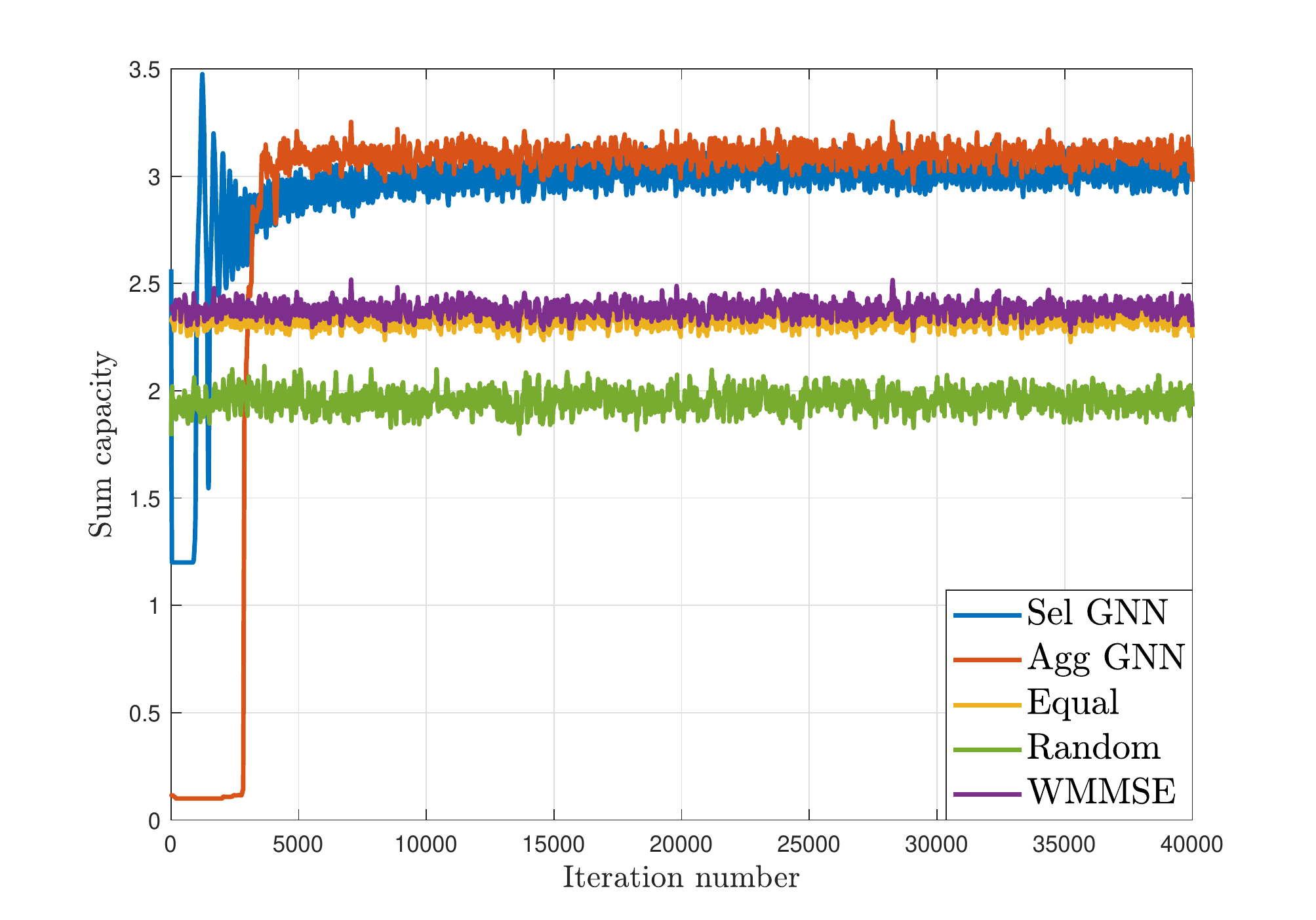}}
\caption{Performance comparison during training for 5 BS and 50 Users. (Hop size=8).}
\label{fig:cell5_10}
\end{figure}

\begin{figure}[t]
 \centering
\includegraphics[width=0.5\textwidth]{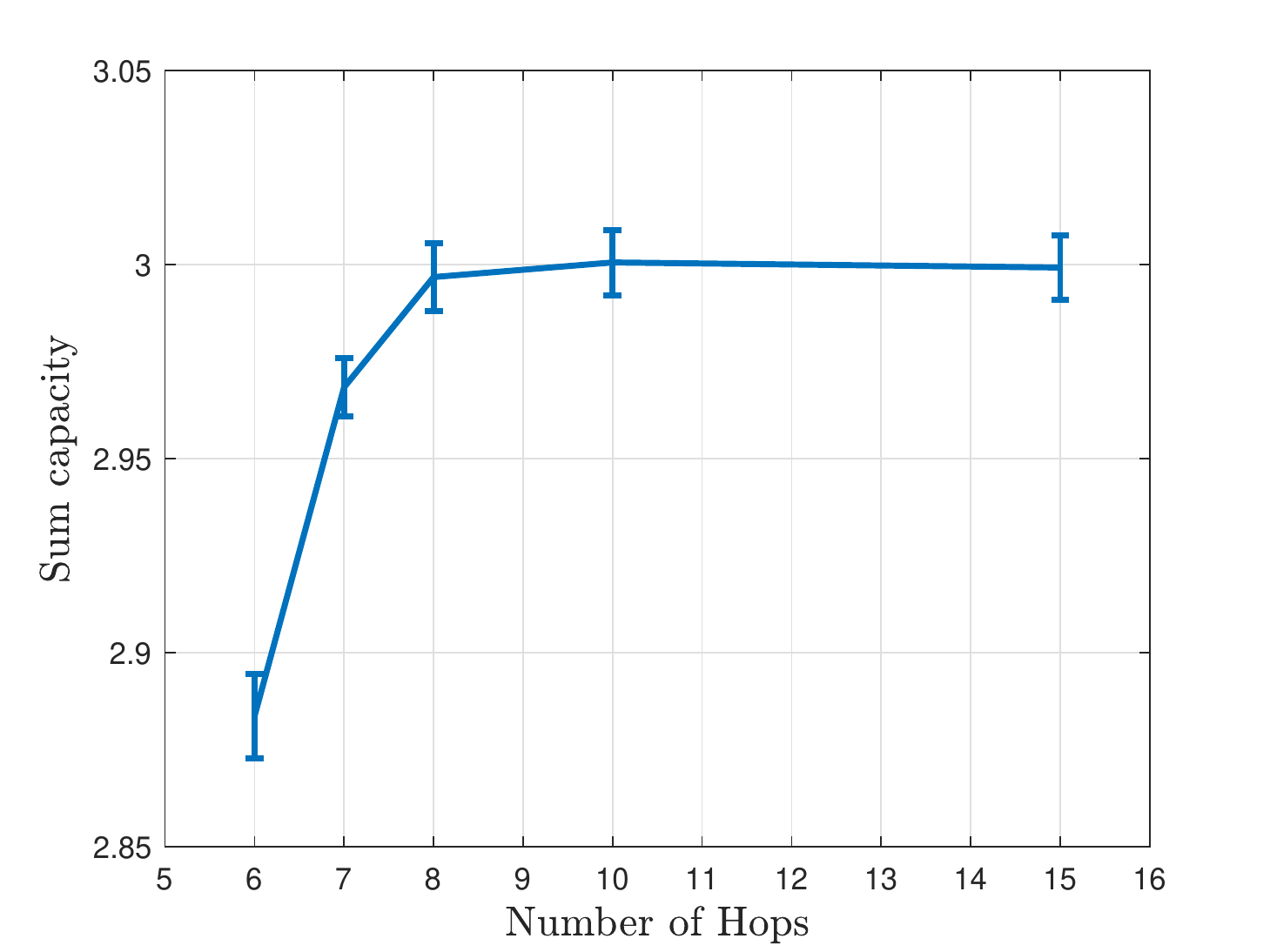}
 \caption{Performance comparison of different hop sizes for 50 users in cellular setting.}
 \label{fig:cellhop}
 \end{figure}

%\blue{expand this}
Lastly, we consider the cellular network scenario where $n$ base stations serve $m$ cellular users which are distributed evenly around the correspondent base station. In Figure \ref{fig:cell5_10}, we show the performance of Agg-GNN compared with other methods during the training process on a network of $n=5$ cells and $m=50$ users. We can see that Agg-GNN still outperforms other methods in this large and practical network setting. In Figure \ref{fig:cellhop}, performance are compared with different values of maximal neighborhood range. We can see that the performance also converges after a certain length. With more channels involved in this cellular setting, the minimum number of information exchanges is also larger compared to the previous adhoc setting. 

% \begin{figure}[t]
% \centering
% \centerline{\includegraphics[width=0.5\textwidth]{figures/cell25.eps}}
% \caption{Performance comparison during training for 5 BS and 25 Users. (Hop size=5).}
% \label{fig:cell5_5}
% \end{figure}

%%%%%%%%%%%%%%%%%%%%%%%%%%%%%%%%%%%%%%%%%%%%%%%%%%%%%%%%%%%%%%%%%%%%%
%%%%%%%%%%%%%%%%%%%%%%%%%%%%%%%%%%%%%%%%%%%%%%%%%%%%%%%%%%%%%%%%%%%%
%%%   S  E  C  T  I  O  N   %%%%%%%%%%%%%%%%%%%%%%%%%%%%%%%%%%%%%%
%%%%%%%%%%%%%%%%%%%%%%%%%%%%%%%%%%%%%%%%%%%%%%%%%%%%%%%%%%%%%%%%%%%%
%%%%%%%%%%%%%%%%%%%%%%%%%%%%%%%%%%%%%%%%%%%%%%%%%%%%%%%%%%%%%%%%%%%%%
\section{Conclusion}
\label{sec:con}
We consider the problem of decentralized resource allocations in wireless networks. By parameterizing the resource allocation function, we can train a graph neural network with primal-dual model-free learning method. Each node can get locally aggregated information from its active neighbors with some delay, which incorporates the underlying network structure of the system. {We consider both synchronous and asynchronous settings by involving heterogeneous working patterns for each node.} We propose a policy based on Aggregation Graph Neural Networks, whose dimension does not scale with network size and can be implemented over the air. We further prove the algorithm preserves the permutation equivariance with respect to the network structure. We verify our results with a series of numerical simulation results demonstrating strong performance and transference.

\appendices
%%%%%%%%%%%%%%%%%%%%%%%%%%%%%%%%%%%%%%%%%%%%%%%%%%%%%%%%%%%%%%%%%%%%%
%%%%%%%%%%%%%%%%%%%%%%%%%%%%%%%%%%%%%%%%%%%%%%%%%%%%%%%%%%%%%%%%%%%%
%%%   S  E  C  T  I  O  N   %%%%%%%%%%%%%%%%%%%%%%%%%%%%%%%%%%%%%%
%%%%%%%%%%%%%%%%%%%%%%%%%%%%%%%%%%%%%%%%%%%%%%%%%%%%%%%%%%%%%%%%%%%%
%%%%%%%%%%%%%%%%%%%%%%%%%%%%%%%%%%%%%%%%%%%%%%%%%%%%%%%%%%%%%%%%%%%%%
\section{Proof of Proposition \ref{prop:permute-gnn}}
\begin{proof}
We first begin with the simple version omitting the time stamps and a constant $\bbH$ matrix. We look at the layer $l=1$ and take the input $\hat\bby_{i0} = [[\hat\bbx]_i, [\hat\bbH\hat\bbx]_i, \hdots, [\hat\bbH^{K-1}\hat\bbx]_i ]$.
At node $i$, the output of the first layer is given by:
%\blue{avoid multi-line equations like this. 2 or 3 lines maximum, but stick to 1 equation when possible. 8 is too many :p In general, you dont have to show every algebraic operation, just the main points}
\begin{align*}
\hat{\bby}_{i1} &=\sigma_1[\bm\alpha_1 * \hat{\bby}_{i0}]\\
    & = \sigma_1 [\bm\alpha_1 * [[\hat{\bbx}]_i; [\hat{\bbH}\hat{\bbx}]_i;\hdots [\hat{\bbH}^{K-1} \hat{\bbx}]_i ]].
\end{align*}
First we realize that for a general $k$, we have:
\begin{align*}
     \hat\bbH^k = \bm\Pi^T\bbH\bm\Pi\bm\Pi^T\bbH\bm\Pi\hdots \bm\Pi^T\bbH\bm\Pi = \bm\Pi^T\bbH^k\bm\Pi,
\end{align*}
due to the fact that $\bm\Pi\bm\Pi^T=\bbI$. By inserting the definition of $\hat\bbx$, we can get
\begin{align*}
  \hat{\bby}_{i1}  & = \sigma_1 [\bm\alpha_1 * [[\bm\Pi^T \bbx]_i; [ \bm\Pi^T \bbH \bbx]_i;\hdots [\bm\Pi^T\bbH^{K-1}  \bbx]_i ]]\\
    & = \sigma_1 [\bm\alpha_1 * [\bm\Pi^T \bbY]_i ]= \sigma_1 [ [\bm\alpha_1 * (\bm\Pi^T \bbY)]_i ],
% \hat{\bby}_{i}^1 &=\sigma^1[\bba^1 * \hat{\bby}_{i}^0]\\
%     & = \sigma^1 [\bba^1 * [[\hat{\bbx}]_i; [\hat{\bbH}\hat{\bbx}]_i;\hdots [\hat{\bbH}^{K-1} \hat{\bbx}]_i ]].
% \end{align*}
% By inserting the definition of $\hat\bbx$ and the matrix definition in \eqref{eqn:matrix}, we can get
% \begin{align*}
%   \hat{\bby}_{i}^1  & = \sigma^1 [\bba^1 * [[\bm\Pi^T \bbx]_i; [ \bm\Pi^T \bbH \bbx]_i;\cdots [\bm\Pi^T\bbH^{K-1}  \bbx]_i ]]\\
%     & = \sigma^1 [\bba^1 * [\bm\Pi^T \bbY]_i ]\\
%     & = \sigma^1 [ [\bba^1 * (\bm\Pi^T \bbY)]_i ].
\end{align*}
where $\bbY(t)$ stands for a matrix representation of the sequences of signals:
\begin{gather}
\label{eqn:matrix}
    \bbY(t):=[\bby^{(0)}(t), \bby^{(1)}(t), \hdots, \bby^{(K-1)}(t)].
\end{gather}
As the convolution and matrix permutation are linear operations, we can get:
\begin{align}
\nonumber\hat{\bby}_{i1}  & = \sigma_1 [[ \bm\Pi^T (\bm\alpha_1 * \bbY)]_i ]= \sigma_1 [\bm\Pi^T[\bm\alpha_1 * \bbY]_i]\\
\label{eqn:permu2}     & = \bm\Pi^T \sigma_1[\bm\alpha_1 * \bby_{i0} ]=\bm\Pi^T \bby_{i1},
%  \nonumber\hat{\bby}_{i}^1  & = \sigma^1 [[ \bm\Pi^T (\bba^1 * \bbY)]_i ]\\
%  \nonumber & = \sigma^1 [\bm\Pi^T[\bba^1 * \bbY]_i]\\
% \label{eqn:permu2}     & = \bm\Pi^T \sigma^1[\bba^1 * \bby_{i}^0 ]=\bm\Pi^T \bby_{i}^1,
    % \sigma[\sum_{n=0}^{K_l-1}a_{ln} [\prod\limits_{m=1}\limits^{K-n+1} \hat{\bbH}_{t_0+m} \hat{\bbx}_{t_0}  ]_i ]\\
    % & = \sigma[\sum_{n=0}^{K_l-1}a_{ln} [\Pi^T   \bbH_{t_0+1}\Pi\Pi^T \bbH_{t_0+2}\cdots \bbH_{t_0+K-n+1}\Pi \Pi^T \bbx_{t_0}  ]_i ]\\
    % & = \sigma[\sum_{n=0}^{K_l-1}a_{ln} [\Pi^T \prod\limits_{m=1}\limits^{K-n+1}  \bbH_{t_0+m} \bbx_{t_0}  ]_i ]\\
    % & = \sigma[\sum_{n=0}^{K_l-1}a_{ln} [\Pi^T \mathbf{z}^i_t(l-1) ]_i]
\end{align}
where $\bby_{i1}$ is the output of the first layer under unpermuted inputs. \eqref{eqn:permu2} is derived based on the pairwise operation $\sigma$. As we have shown the output of a single layer is permutation equivalent to its input, it can be concluded that the output of layers $l=2,3,\hdots, L$ is also permutation equivalent. 

Moreover, the exponent of the constant $\bbH$ matrix can be replaced with a product sequence of time varying $\bbH(t)$ matrix with the equalities still hold, i.e.
\begin{gather}
\prod\limits_{i=1}^{k-1}\hat\bbH(t-i)\hat\bbx(t-k)=\bm\Pi^T\prod\limits_{i=1}^{k-1}\bbH(t-i)\bbx(t-k).
\end{gather}

In addition, the input history information we consider is actually a limited channel matrix represented by \eqref{eqn:limitchannel}. With $\hat\bbQ=\bm\Pi^T \bbQ \bm\Pi$, it can be derived that:
\begin{gather}
 \hat\bbH(t) \circ \hat\bbQ(t) = \bm\Pi^T \bbH(t) \bm\Pi \circ \bm\Pi^T \bbQ(t) \bm\Pi =\bm\Pi^T (\bbH(t)\circ\bbQ(t)) \bm\Pi.
\end{gather}
Based on our previous setting, $\bbH(t)\circ\bbQ(t)=\tbH(t)$ can be seen as an equivalent channel matrix and the above derivation process still holds. This therefore comes to the conclusion that $\bm\Phi(\hat\ccalH,\bbA)=\bm\Pi^T \bm\Phi(\ccalH,\bbA)$.
\end{proof}

%%%%%%%%%%%%%%%%%%%%%%%%%%%%%%%%%%%%%%%%%%%%%%%%%%%%%%%%%%%%%%%%%%%%%
%%%%%%%%%%%%%%%%%%%%%%%%%%%%%%%%%%%%%%%%%%%%%%%%%%%%%%%%%%%%%%%%%%%%
%%%   S  E  C  T  I  O  N   %%%%%%%%%%%%%%%%%%%%%%%%%%%%%%%%%%%%%%
%%%%%%%%%%%%%%%%%%%%%%%%%%%%%%%%%%%%%%%%%%%%%%%%%%%%%%%%%%%%%%%%%%%%
%%%%%%%%%%%%%%%%%%%%%%%%%%%%%%%%%%%%%%%%%%%%%%%%%%%%%%%%%%%%%%%%%%%%%
\section{Proof of Proposition \ref{prop:permute-policy}}
\begin{proof}
To prove \eqref{eqn:u} holds, we need to first prove that $\hat\bbr=\bm\Pi^T \bbr$. To begin with, we have:
\begin{gather}
\label{eqn:prop2-1} 
    \hat\bbr=\int \bbf\left( \hat\bbP(\hat\ccalH), \hat\bbH,\hat\bbx \right)\text{d}\hat{m}(\hat\bbH,\hat\bbx).
\end{gather}
Combining with the assumption that $\hat\bbP(\hat\ccalH)=\bm\Pi^T \bbP(\ccalH)$, we can get
\begin{gather}
    \hat\bbr=\int \bbf\left( \bm\Pi^T \bbP(\ccalH),\hat\bbH, \hat\bbx\right) \text{d}\hat{m}(\hat\bbH,\hat\bbx).
\end{gather}
Implement the change of variables of $\hat\bbH$ and $\hat\bbx$ to get
\begin{align}
    \hat\bbr=\int \bbf \left( \bm\Pi^T \bbP(\ccalH), \bm\Pi^T \bbH \bm\Pi, \bm\Pi^T\bbx\right)\text{d}\hat{m}(\bm\Pi^T\bbH\bm\Pi, \bm\Pi^T\bbx).
\end{align}
With the assumption that function $\bbf$ is permutation equivariant, there is $\bbf\left( \bm\Pi^T \bbP(\ccalH), \bm\Pi^T \bbH \bm\Pi, \bm\Pi^T \bbx\right) = \bm\Pi^T \bbf(\bbP(\ccalH), \bbH,\bbx)$. Together with \eqref{eqn:Hpermute}, this leads to:
\begin{align}
    \hat\bbr=\int \bm\Pi^T \bbf\left( \bbP(\ccalH), \bbH,\bbx \right)\text{d} m(\bbH,\bbx).
\end{align}
Taking the permutation matrix out of the integration we can get
\begin{align}
\label{eqn:permute_r}
    \hat\bbr =\bm\Pi^T \bbr.
\end{align}
With permutation invariance assumptions of $u_0(\bbr)$ and $\bbu(\bbr)$, result \eqref{eqn:u} can be derived directly. Therefore, the permutations of the network and its correspondent resource allocation functions result in rewards with the same utility, which indicates \eqref{eqn:P} holds.
\end{proof}

\section{Proof of Theorem \ref{thm:trans}}
\begin{proof}
For problem \eqref{eqn:opt}, suppose that we have the optimal solution for distribution $m(\bbH, \bbx)$ as $\bbA^*$ with the optimal strategy and reward denoted as $\bm\Phi(\ccalH,\bbA^*)$ and $\bbr^*$ respectively. Permute the network with matrix $\bm\Pi\in\psi$, based on Proposition \ref{prop:permute-gnn}, the parameterized allocation strategy satisfies:
\begin{align}
    \bm\Phi(\hat\ccalH,\bbA^*)=\bm\Pi^T\bm\Phi(\ccalH,\bbA^*).
\end{align}
Following the derivation from \eqref{eqn:prop2-1} to \eqref{eqn:permute_r}, 
\begin{align}
    \hat\bbr&=\int \bbf\left( \bm\Phi(\hat\ccalH,\bbA^*), \hat\bbH,\hat\bbx \right)\text{d}\hat{m}(\hat\bbH,\hat\bbx)\\
    &=\bm\Pi^T\bbr =\int \bbf\left( \bm\Phi( \ccalH,\bbA^*),  \bbH, \bbx \right)\text{d} {m}(\bbH, \bbx).
\end{align}
With the permutation invariance of $u_0$, we have $\bbu(\bbr)=\bbu(\hat\bbr)$.

Similarly, we suppose the optimal solution for distribution $\hat m(\hat\bbH, \hat\bbx)$ is $\hat\bbA^*$ and we have that:
\begin{align}
    &\bm\Phi(\hat\ccalH,\hat\bbA^*)=\bm\Pi^T\bm\Phi(\ccalH,\hat\bbA^*).\\
     \hat\bbr^*&=\int \bbf\left( \bm\Phi( \hat\ccalH,\hat\bbA^*),  \hat\bbH, \hat\bbx \right)\text{d}\hat{m}( \hat\bbH, \hat\bbx)\\
    &=\bm\Pi^T\bbr =\int \bbf\left( \bm\Phi( \ccalH,\hat\bbA^*),  \bbH, \bbx \right)\text{d} {m}(\bbH, \bbx).
\end{align}
The utility function here satisfies $\bbu(\bbr)=\bbu(\hat\bbr^*)$. $\hat\bbr^*$ is optimal for $\hat m(\hat\bbH, \hat\bbx)$ and $\bbr$ is feasible, while  $\bbr^*$ is optimal for $m(\bbH, \bbx)$ and $\bbr$ is feasible, we have 
\begin{align}
    u_0(\hat\bbr^*)\geq u_0(\hat\bbr)=u_0(\bbr^*), u_0(\bbr^*)\geq u_0(\bbr)=u(\bbr^*).
\end{align}
Therefore, the inequalities must be equalities, which means $\bbA^*$ is optimal for $\hat m(\hat\bbH, \hat\bbx)$ and $\hat\bbA^*$ is optimal for $m(\bbH, \bbx)$. This concludes the proof.
\end{proof}

\urlstyle{same}
\bibliographystyle{IEEEtran}
\bibliography{references}
\end{document}